\documentclass[english,11pt,a4paper]{article}
\usepackage{authblk}
\usepackage[text={17.0cm,23.7cm}]{geometry}
\usepackage[T1]{fontenc}
\usepackage[utf8]{inputenc}
\usepackage{babel}
\usepackage{graphicx}
\usepackage{tabularx}
\usepackage{amsmath}
\usepackage{multirow}
\usepackage{booktabs}
\usepackage{amssymb}
\usepackage[dvipsnames]{xcolor}
\usepackage{yfonts}
\usepackage{ulem}
\usepackage{enumitem}
\usepackage{mdframed}
\usepackage{xcolor}
\usepackage[style=numeric-comp, sorting=none]{biblatex}
\usepackage{float}
\usepackage{mathrsfs}

\usepackage{biblatex}
\usepackage[colorlinks=true, allcolors=blue]{hyperref}

\begin{document}
	
	\title{\vspace{-2cm}

		\vspace{0.6cm}

	\textbf{A neutrino floor for the Migdal effect}\\[8mm]}

	\author[1,2,3]{Gonzalo Herrera}
    \affil[1]{\normalsize\textit{Center for Neutrino Physics, Department of Physics, Virginia Tech, Blacksburg, VA 24061, USA}}
	\affil[2]{\normalsize\textit{Physik-Department, Technische Universit\"at M\"unchen, James-Franck-Stra\ss{}e, 85748 Garching, Germany}}
	\affil[3]{\normalsize\textit{Max-Planck-Institut f\"ur Physik (Werner-Heisenberg-Institut), F\"ohringer Ring 6, 80805 M\"unchen, Germany}}
	\date{}
	
	\maketitle
	
	\begin{abstract}
    Neutrino-nucleus scatterings in the detector could induce electron ionization signatures due to the Migdal effect. We derive prospects for a future detection of the Migdal effect via coherent elastic solar neutrino-nucleus scatterings in liquid xenon detectors, and discuss the irreducible background that it constitutes for the Migdal effect caused by light dark matter-nucleus scatterings. Furthermore, we explore the ionization signal induced by some neutrino electromagnetic and non-standard interactions on nuclei. In certain scenarios, we find a distinct peak on the ionization spectrum of xenon around 0.1 keV, in clear contrast to the Standard Model expectation.
	\end{abstract}

\section{Introduction}

In some problems in perturbation theory, the time of action of the perturbation can be small, while the perturbation itself does not need to be weak. One example of this circumstance is the ionization due to $\beta$ decay. The velocity of the outgoing electron is much larger than the velocity of the atomic electrons, thus, from the reference frame of the atomic electrons, the change in the nuclear charge occurs suddenly. The sudden approximation in Quantum Mechanics then allows to compute transition probabilities in systems whose wavefunction is not substantially changed, $\Phi(t) \approx \Phi(0)$.

One interesting application of the sudden approximation was proposed by Migdal in 1939 \cite{migdal:1939svj}, later discussed in his textbook in Quantum Mechanics \cite{Migdal1977QualitativeMI}. In nuclear collisions involving large energy transfers, ionization of the recoil atoms is likely to occur. If the velocity of the nucleus is small, then it carries its electrons off with it. For large velocities, on the other hand, the nucleus recoils out of its electronic shells and electrons do not have time to catch up with the nucleus motion. Evidence for the Migdal effect has been reported in measurements of compton scattering of neutrons in hydrogen \cite{compton_anomaly}, and in some nuclear decay processes \cite{PhysRevLett.108.243201}. Further, there are some experimental proposals aiming to observe the Migdal effect from nuclear recoils of Standard Model particles in the near future \cite{Nakamura:2020kex,Bell:2021ihi,Araujo:2022wjh}. A first dedicated search in liquid xenon from neutron scatters was already performed, albeit finding no signal consistent with the model expectations \cite{Xu:2023wev}. This may indicate that the theoretical uncertainties are large and difficult to quantify \cite{Liu:2020pat, Baxter:2019pnz}.

The importance of the contribution from the Migdal effect for direct dark matter searches was initially discussed in \cite{Vergados:2005dpd,Moustakidis:2005gx,Bernabei:2007jz}, and several subsequent studies have followed in recent years, \textit{E.g} \cite{Ibe:2017yqa, Dolan:2017xbu, Essig:2019xkx, Bell:2019egg, Baxter:2019pnz, Knapen:2020aky, Liu:2020pat, GrillidiCortona:2020owp,Tomar:2022ofh, Wang:2021oha, Berghaus:2022pbu, Adams:2022zvg, Cox:2022ekg, Blanco:2022pkt, Li:2023xkf, Bell:2023uvf, Gu:2023pfg, AtzoriCorona:2023ais}. The ionization signal from the Migdal effect is suppressed w.r.t the nuclear one, but there is a range of energies where the nuclear recoil energy falls below threshold while the electromagnetic signal is observable.

Moreover, it has been pointed out for some years that direct detection experiments, originally designed to detect scatterings of dark matter particles from the Milky Way halo with the detector, could also be used to study solar neutrino physics (in the Standard Model and beyond), \textit{E.g} \cite{Pospelov:2011ha, Harnik:2012ni, Billard:2014yka, Dent:2016wcr, Cerdeno:2016sfi, Dutta:2017nht, AristizabalSierra:2017joc, Boehm:2018sux}. Crucially, it has been claimed that scatterings of solar neutrinos off nuclei in the Standard Model may constitute an irreducible background for direct dark matter searches, the so-called neutrino "floor", or neutrino "fog" \cite{Monroe:2007xp,Vergados:2008jp,Gutlein:2010tq,OHare:2021utq}. 

In this work, we aim to study the current possibilities and future prospects to detect the Migdal effect electronic ionization induced by coherent elastic neutrino-nucleus scattering at Earth-based liquid xenon detectors, and its distinguishability from a light dark matter induced signal.

In particular, we consider in detail the full low-energy neutrino flux at Earth stemming from processes in the Sun such as the proton-electron-proton reaction ($pep$), the $hep$ branch, the  $^{7}$Be line, the $\beta^{+}$ decay processes,  $^{13}$N, $^{15}$O and $^{17}$F, as well as the contribution from the diffuse supernova neutrino background and atmospheric neutrinos \cite{Billard:2013qya, Vitagliano:2019yzm}. It should be noted that previous works calculated the Migdal ionization rate induced by solar neutrinos from $pp$, $^{8}$B, $pep$ and $^{7}$Be processes, plus the atmospheric neutrino flux \cite{Ibe:2017yqa, Bell:2019egg}. Most of the additional neutrino fluxes that we include in our calculation are largely subdominant with respect to these, with the exception of the  $^{13}$N, $^{15}$O and $^{17}$F processes, which dominate in certain energy regions between $\sim 420$ keV and $\sim 1800$ keV, in between the $pp$ and $^{8}B$ fluxes \cite{Billard:2014yka}. Further, we will confront our results with XENON1T (S2-only) data \cite{Aprile_2019}, and with the expected signal from light dark matter-electron scatterings, in order to discuss its discriminability. Additionally, we will study the ionization signal induced by solar neutrino scatterings off nuclei due to electromagnetic and non-standard interactions.

The paper is organized as follows: In section \ref{sec:Migdal_rate_sevens_DM}, we will introduce the relevant formalism and calculate the Migdal ionization rate induced by CE$\nu$NS in liquid xenon, further confronting it with the signal induced by light dark matter-nucleus scatterings, and deriving the neutrino "floor" induced by the Migdal effect. Further, we derive projections on the future detectability of the Migdal effect induced by solar neutrino-nucleus scatterings in XENONnT. In section \ref{sec:ionization_signal_NSI}, we will calculate the ionization rate induced by electromagnetic and some non-standard neutrino-nucleus interactions, noticing a distinct feature on the ionization spectrum around 0.1 keV in some beyond the Standard Model scenarios arising from the ionization of electrons in the $n=4$ shell by $pp$-chain solar neutrinos. Finally, in section \ref{sec:conclusions}, we will present our conclusions.

\section{Migdal ionization rate due to CE$\nu$NS and dark matter-nucleus scatterings}\label{sec:Migdal_rate_sevens_DM}
The CE$\nu$NS process can induce ionization through the Migdal effect. The differential cross-section for this process is \cite{Freedman:1973yd,Migdal1977QualitativeMI,Ibe:2017yqa, Bell:2019egg}

\begin{equation}\label{eq:cenuNs}
    \frac{d\sigma_{\nu N-\nu N^{*} e^{-}}}{d E_R}=\frac{G_F^2 m_A}{4\pi}Q_V^{2}\Big (1-\frac{m_A E_R}{2 E_{\nu}^{2}}\Big ) \times \left |Z_{ion}(E_{er})  \right |^{2}
\end{equation}
where $m_A$ is the target mass. $Q_V$ is the vector charge of the nucleus and reads
\begin{equation}
Q_V=g_{pV} Z F_p\left(E_R\right)+g_{nV} N F_n\left(E_R\right)
\end{equation}
and $F_{p}$ and $F_n$ are the proton and neutron form factor, respectively, for which we adopt the Helm prescription with proton and neutron root mean square radius of and $R_p=3.14$ fm and $R_n=3.36$ fm. $Z$ and $N$ are the number of protons and neutrons in the nucleus, and the neutral current vector couplings, $g_{pV}$ and $g_{nV}$ are

\begin{equation}
g_{pV} =\frac{1}{2}-2 \sin ^2 \theta_W
\end{equation}
\begin{equation}
g_{nV} =-\frac{1}{2}
\end{equation}
where the weak mixing angle is taken as $\sin ^2 \theta_W \simeq 0.237$. The ionization form factor $ \left |Z_{ion}(E_{er})  \right |^{2}$ is given by
\begin{equation}
 \left |Z_{ion}(E_{er})  \right |^{2}=\frac{1}{2\pi}\sum_{n,l}\int dE_{er}\frac{d}{d E_{er}}p(nl \rightarrow E_{er}) 
\end{equation}
where the ionization probability of an electron in the orbital $(n,l)$ is denoted by $p$. We take the tabulated values of the ionization probabilities for xenon from \cite{Ibe:2017yqa}, accounting for the orbitals $5p$, $5s$, $4d$, $4p$, $4s$, $3d$, $3p$ and $3s$. $E_{er}$ refers to the electron recoiling energy. The total ionization rate due to the Migdal effect can then be calculated as

\begin{equation}\label{eq:DM-rate}
\frac{d R^{\rm mig}}{d E_R}=N_T \int_{E_v^{\min }}^{E_v^{\max }} \frac{d \Phi}{d E_v} \frac{d \sigma_{\nu N \rightarrow \nu N^{*}e^{-}}}{d E_R} d E_v
\end{equation}
where $\frac{d \Phi}{d E_v}$ is the differential neutrino flux. We take all known solar, atmospheric and diffuse supernova background contributions from \cite{Billard:2013qya}. $N_T$ is the number of target nuclei in the detector. Finally, the detector energy $E_{\rm det}$ measured at liquid xenon experiments like XENONnT is obtained by convolving the integration of the nuclear recoil rates with $\delta(E_{\rm det}-q_{nr}E_R-E_{er}-E^{nl})$, where $E_R$ is the nuclear recoil energy, $E_{er}$ is the ionized electron recoiling energy, $E^{nl}$ is the orbital binding energy and $q_{nr}$ is the conversion factor from nuclear to electron equivalent energy. We take $q_{nr}=0.15$ \cite{XENON:2019zpr}. The limits of integration over nuclear recoil energies are given by
\begin{equation}
E_R^{\rm min}=\frac{\left(E_{er}+E^{nl}\right)^2}{2 m_A}
\end{equation}
\begin{equation}
E_R^{\rm max}=\frac{\left(2 E_\nu-\left(E_{er}+E^{nl}\right)\right)^2}{2\left(m_A+2 E_\nu\right)}.
\end{equation}

Dark matter particles can also induce ionization via Migdal effect in nuclear scatterings. The rate of ionization events due to the Migdal effect corresponds to the standard dark matter-nucleus differential recoil rate multiplied by the ionization rate
\begin{equation}
    \frac{d^{2}R^{\rm mig}}{dE_{R} dv}= \frac{d^{2}R}{d E_R dv}\times \left |Z_{ion}(E_{er})  \right |^{2}
\end{equation}
where, for a given nucleus $i$
\begin{equation}
\frac{d^{2}R}{d E_R dv}=\mathcal{F}(\vec{v})\frac{\mathrm{d} \sigma_i}{\mathrm{~d} E_R}\left(v, E_R\right)
\end{equation}
where $\mathcal{F}(\vec{v})$ is the total dark matter flux reaching the Earth. We assume the so called Standard Halo Model (SHM), with a local dark matter density of $\rho_{\rm DM}=0.3$ GeV/cm$^{3}$ and a Maxwellian velocity distribution truncated at the escape velocity of the Milky Way, such that
\begin{equation}
\mathcal{F}(\vec{v})=\frac{\rho_{\rm DM}}{m_{\mathrm{DM}}} v f_{\mathrm{MB}}(\vec{v})
\end{equation}
where the Maxwell-Boltzmann velocity distribution in the galactic frame is defined as
\begin{equation}
f_{\mathrm{MB}}(\vec{v})=\frac{1}{\left(2 \pi \sigma_v^2\right)^{3 / 2} N_{\mathrm{esc}}} \exp \left[-\frac{v^2}{2 \sigma_v^2}\right] \quad \text { for } v \leq v_{\mathrm{esc}} \text {, }
\end{equation}
and here, $v=|\vec{v}|, \sigma_v=156 \mathrm{~km} / \mathrm{s}$ is the velocity dispersion \cite{Green:2011bv}, and $v_{\text {esc }}=544 \mathrm{~km} / \mathrm{s}$ is the escape velocity from our galaxy \cite{Deason_2019}. $N_{\text {esc }}$ is a normalization constant with value $N_{\text {esc }} \simeq 0.993$. We point out that further contributions to the total dark matter flux on Earth and corrections to the SHM are expected, \textit{E.g} \cite{Bozorgnia:2016ogo,Necib:2018igl,Evans:2018bqy,Besla:2019xbx,Herrera:2021puj}, but an analysis on the impact of these on the Migdal ionization rate is beyond the scope of our work, and left for future studies. 

For the differential dark matter-nucleus scattering cross section, we assume for simplicity an isoscalar spin-independent dark matter-nucleus interaction, with equal couplings of dark matter to all nucleons
\begin{equation}
\frac{d \sigma_i^{\mathrm{SI}}}{d E_R}\left(v, E_R\right)=\frac{m_{A_i}}{2 \mu_{\rm DM-n}^2 v^2} A_i^2 \sigma_{\rm DM-n} F_i^2\left(E_R\right)
\end{equation}
where $\sigma_{\rm DM-n}$ is the non-relativistic dark matter-nucleon scattering cross section
\begin{figure}[t!]
    \includegraphics[width=0.49\linewidth]{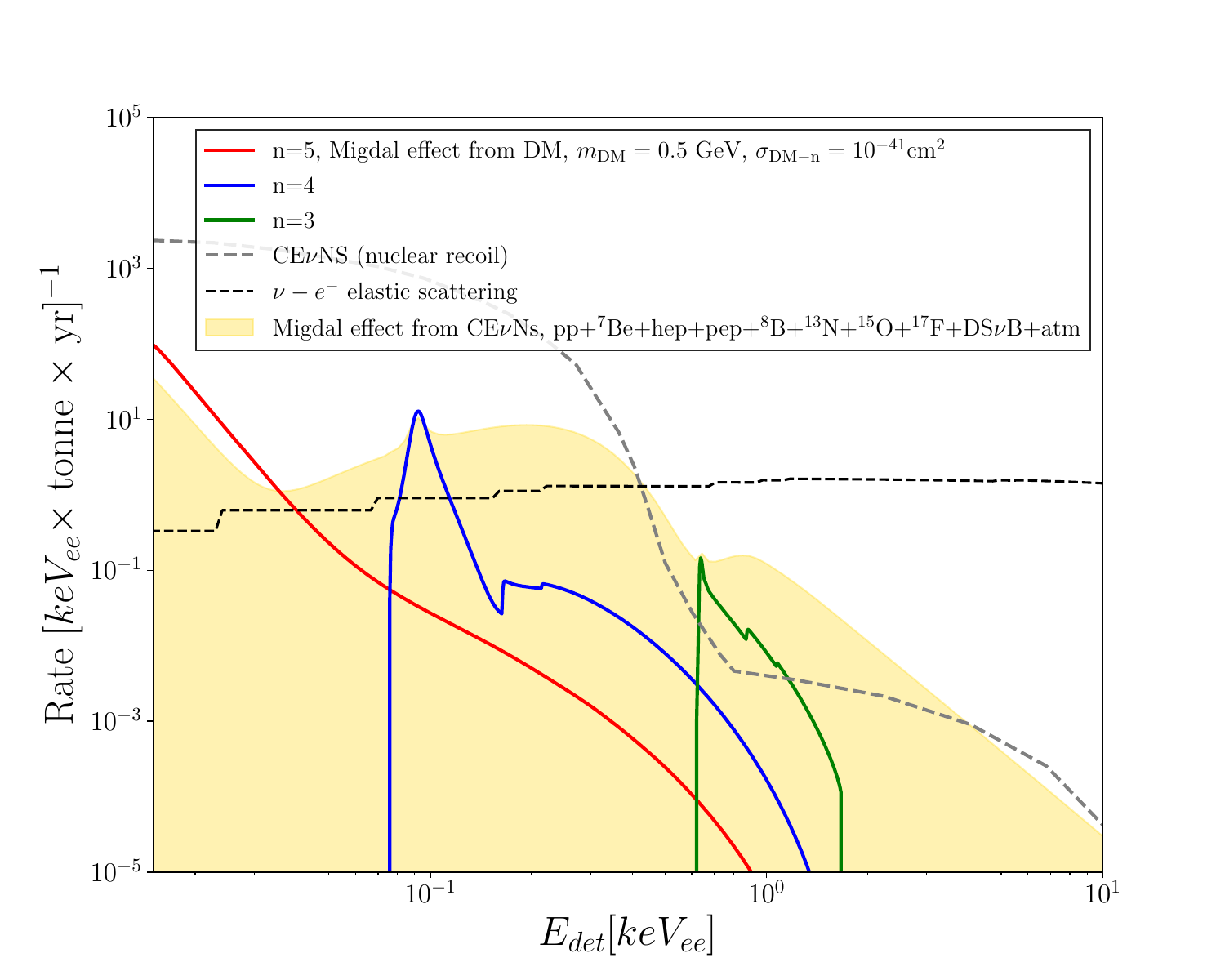}
    \includegraphics[width=0.49\linewidth]{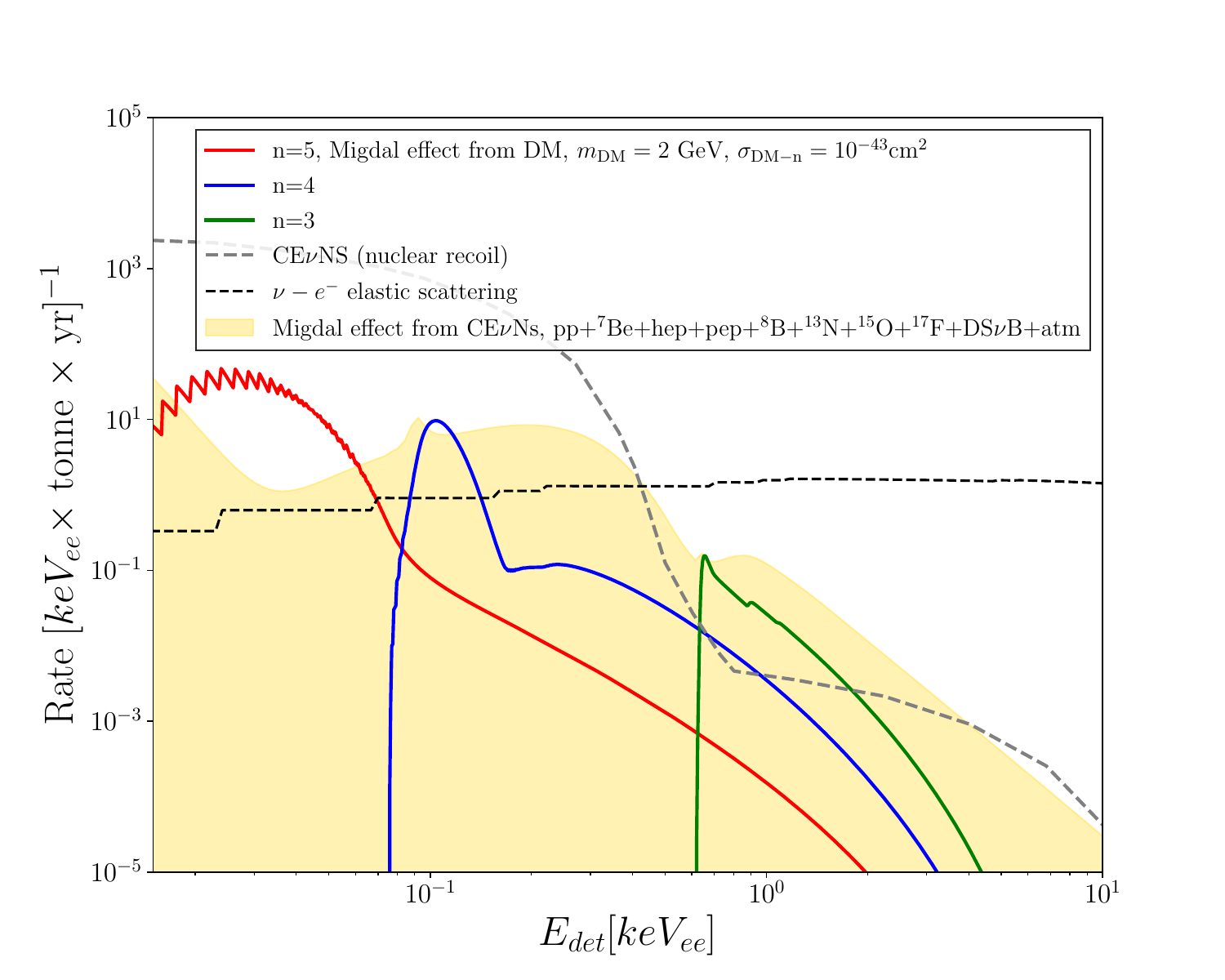}
\caption{\textit{Left plot}: Comparison of the Migdal ionization rate induced by CE$\nu$NS in xenon (yellow) and the Migdal ionization rate induced by a light dark matter particle with mass of $m_{\rm DM}=0.5$ GeV and cross section with nucleons $\sigma_{\rm DM-n}=10^{-41}$ cm$^2$, where we separate the contribution from the shells $n=5$ (red), $n=4$ (blue) and $n=3$ (green). For comparison, we show the nuclear recoil signal from CE$\nu$NS (grey), and the electron ionization signal from elastic neutrino-electron scatterings (black). \textit{Right plot}: Similar plot, for a heavier dark matter particle of mass $m_{\rm DM}=2$ GeV and $\sigma_{\rm DM-n}=10^{-43}$ cm$^{2}$.}
\label{fig:Migdal_sevens_vs_DM}
\end{figure}

With this, after integrating Equation \ref{eq:DM-rate} over the velocity distribution of the dark matter particles, we can calculate the ionization rate as a function of the detector energy. We implement our own code for all these tasks, finding good agreement with previous literature on the topic, \textit{E.g} \cite{Ibe:2017yqa}. In Figure \ref{fig:Migdal_sevens_vs_DM} we show the rate of events in xenon in tonnes $\times$ year from the Migdal ionization rate from CE$\nu$Ns, including the full neutrino flux at Earth at MeV energies. We compare this with the nuclear recoil signal from CE$\nu$Ns, and the electron ionization signal from neutrino-electron scattering. The Migdal ionization signal is expected to dominate over the neutrino-electron scattering signal below $\sim 0.4$ keV$_{ee}$, and over the nuclear recoil signal from $\sim 0.4-4$ keV$_{ee}$. Further, we show the signal rate induced by the Migdal effect from nuclear scatterings of sub-GeV dark matter particles, including the contributions from the outer shells, for different values of the dark matter mass and scattering cross section with nucleons. As apparent in the plot, for such values, the irreducible background from solar neutrinos would mask the signal induced by light dark matter inducing ionizations via nuclear scatterings and the Migdal effect. This indicates that the neutrino floor for the Migdal effect induced by light dark matter-nucleon scatterings in xenon-based detectors can be present already for cross sections of $\sigma_{\rm DM-p} \sim 10^{-42}$ cm$^2$. Current upper limits on the Migdal effect from light dark matter-nucleon scattering constrain values of $\sigma_{\rm DM-p} \sim 10^{-38}$ cm$^2$ at $m_{\rm DM} \sim 0.5$ GeV \cite{Aprile_2019}, so there is yet roughly 4 orders of magnitude in sensitivity to explore before needing to fit accurately the neutrino floor for the Migdal effect, in order to claim a potential detection of light dark matter particles in the future.

This qualitative discussion from Figure \ref{fig:Migdal_sevens_vs_DM} can be extended more quantitatively by deriving the neutrino floor for the Migdal effect signal from dark matter, in the parameter space of dark matter mass vs scattering cross section. As a criteria to find the values of the dark matter-nucleon scattering cross section corresponding to the neutrino floor, we impose that the ionization rate from CE$\nu$Ns should exceed the ionization rate induced by dark matter over the relevant electron recoil energy range of xenon-based experiments. Under this criteria, the dark matter signal would be "masked" by the neutrino signal, however, we remind that with sufficient statistics and via spectral analyses, both contributions might be distinguishable. Indeed, it has been long discussed that the neutrino floor should be rather regarded as the neutrino "fog".

\begin{figure}[t!]
    \includegraphics[width=0.49\linewidth]{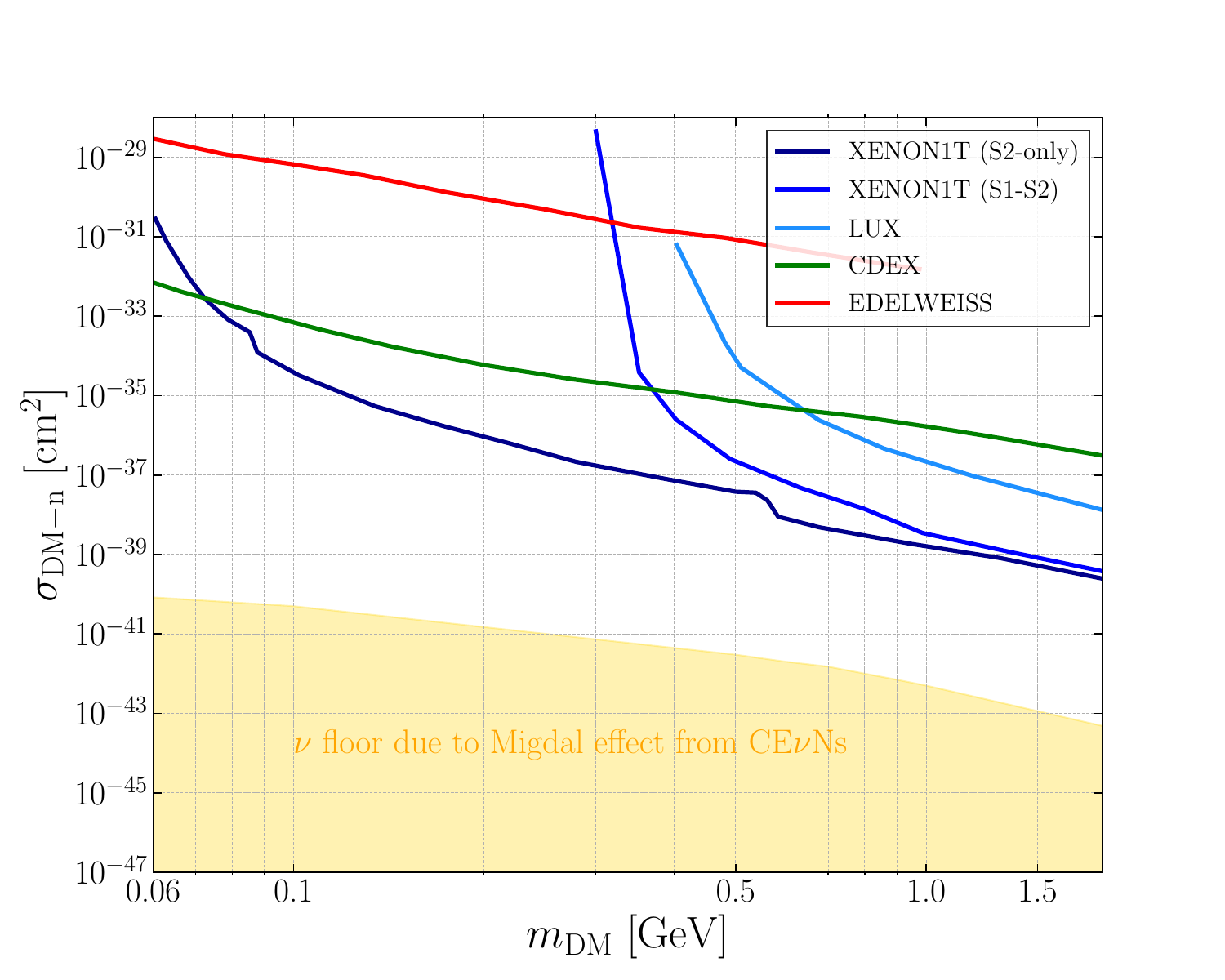}
    \includegraphics[width=0.49\linewidth]{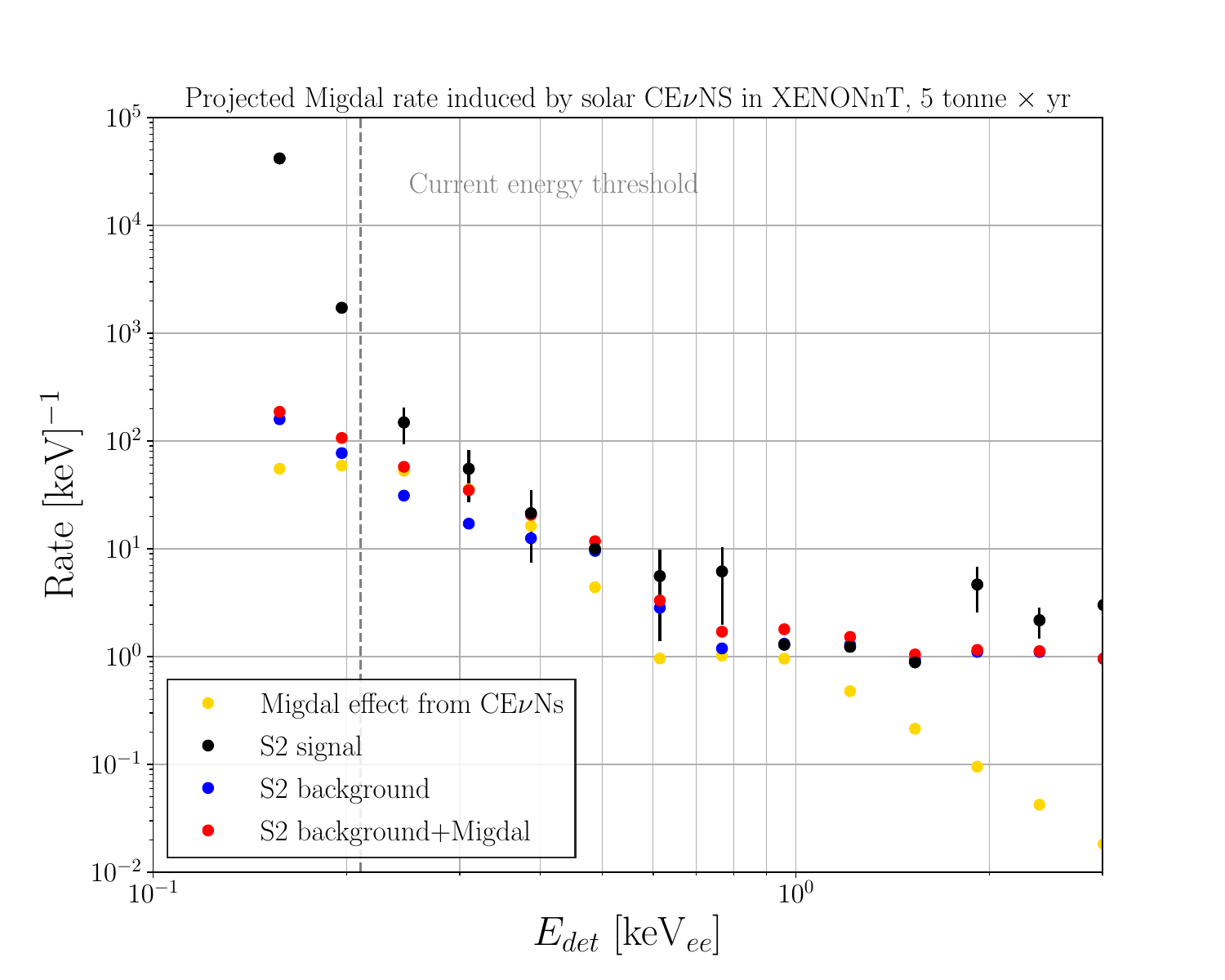}
    \caption{\textit{Left plot}: Neutrino floor in Xenon due to the ionization of electrons following the coherent elastic neutrino-nucleus scattering. For comparison, we show current upper limits on the light dark matter-nucleon scattering cross section from the Migdal effect in XENON1T \cite{XENON:2019zpr}, LUX \cite{LUX:2018akb}, CDEX \cite{CDEX:2019hzn}, and EDELWEISS \cite{Armengaud_2022}. \textit{Right plot}: Projected rate of Migdal ionization events in XENONnT from the CE$\nu$Ns, and its distinguisability from the background events. See main text for further details.}
\label{fig:Migdal_nufloor}
\end{figure}

In the left plot of Figure \ref{fig:Migdal_nufloor} we derive the neutrino floor induced by the Migdal ionization signal from CE$\nu$NS in Xenon, and confront it with upper limits derived from the Migdal effect in XENON1T (S2 only and S1-S2 analyses), LUX, CDEX and the EDELWEISS experiments. As can be appreciated in the Figure, for dark matter masses between $m_{\rm DM} \sim 0.5-2$ GeV, current sensitivity of XENON1T allows to explore less than four orders of magnitude in the values of the dark matter-nucleon scattering cross section, before the neutrino floor is reached. For dark matter masses between $m_{\rm DM} \sim 0.08-0.5$ GeV, this range extends from 4 to 5 orders of magnitude, and for even lower dark matter masses, the neutrino floor is as far as 7 orders of magnitude from current sensitivity.

Given the substantial increase in sensitivity at XENONnT, LUX-ZEPLIN and PANDA-X most recent runs \cite{XENON:2023cxc, LZ:2022lsv, PandaX-II:2021nsg}, it is plausible to speculate that dedicated analysis on the Migdal effect with near future data may allow to detect the ionization contribution from the Migdal effect due to CE$\nu$NS. In the following, we calculate the projected impact of ionizations due to Migdal effect in the XENONnT experiment.

The lowest electron recoiling energies reached by XENON are found in the S2-only data analysis \cite{Aprile_2019}, with energy threshold as small as 0.2 keV$_{ee}$. At slightly lower energies, the observed rate is significantly larger than the background prediction, but the collaboration claims no excess due to lack of knowledge on the background at such low energies, and due to their small efficiency. In the following, we study directly the detectability of the Migdal effect from CE$\nu$NS for current exposures from XENONnT experiment \cite{XENON:2022mpc}, and assuming a similar background and signal per tonne $\times$ year as XENONnT analysis \cite{XENON:2023cxc}, and same efficiency function as in XENON S2-only analysis \cite{Aprile_2019}. The most recent XENONnT analysis of electron recoil data achieved an astonishing low background level in the 1-30 keV region of $15.8 \pm 1.3$ events per tonne $\times$ year $\times$ keV. We will consider this background level for our projection, assuming it might be preserved at the lower recoil energies considered in previous S2-only data analysis. Furthermore, we neglect the neutrino-electron scattering signal in our analysis, which is expected to dominate over the Migdal induced signal at energies above 1 keV, but is subdominant by 1 to 2 orders of magnitude in the energy range from $\sim 0.01 to \sim 0.4$ keV, where the Migdal signal from neutrinos might be detected.

In the right panel of Figure \ref{fig:Migdal_nufloor}, we show our results. The Migdal effect induced by CE$\nu$Ns is expected to dominate over current low-energy backgrounds for a XENON1T S2-like data in the energy range from 0.1 to 1 keV$_{ee}$ for exposures of $\sim$ tonnes $\times$ year, and background levels between 1 to 10 events per $\sim$ keV $\times$ year $\times$ tonne, as was recently achieved in the energy range from 1 to 10 keV in XENONnT \cite{Aprile:2022vux}. Thus, the contribution from the Migdal effect is expected to become sizable in the future. At larger energies than 1 keV$_{ee}$, the sensitivity reach would need to be improved by orders of magnitude. Lowering threshold and background levels at low-energies will be crucial for the detection of the Migdal effect from neutrinos. This, in turn, is a necessary task to be able to detect the Migdal effect induced by light dark matter scatterings off nuclei in the detector. It should be noted, however, that in our analysis we are assuming a future data set analogous to the most recent XENON1T S2-only data, but in reality we are assuming that the nuclear recoil can be told apart from the electron recoil signal, either via a combined S1-S2 analysis, or via a dedicated spectral analysis, that could differentiate the nuclear from the Migdal ionization signals. Also, the current energy threshold of the XENON1T collaboration for the combined S1-S2 data is not as small as that in the XENON1T S2-only analysis, so our projections will only apply if the energy threshold for a combined S1-S2 analysis is reduced in the future.

\section{The ionization signal induced by electromagnetic and non-standard neutrino-nucleus scatterings}\label{sec:ionization_signal_NSI}
Neutrinos are neutral particles in the Standard Model, but are guaranteed to interact electromagnetically at one-loop with nuclei via multipoles. The number of this multipoles allowed and whether they are diagonal or off-diagonal depends on the Majorana or Dirac nature of the neutrino \cite{Nieves,Kayser,Giunti:2014ixa}.

If neutrinos scatter coherently on nuclei via a magnetic moment, the associated ionization due to the Migdal effect has cross section \cite{Vogel:1989iv}

\begin{equation}
\left(\frac{\mathrm{d}\sigma_{\nu N \rightarrow \nu N^{*} e^{-}}}{\mathrm{d} E_R}\right)_{\mu}=\frac{\pi \alpha^2 \mu_{\nu, \rm eff}^2 Z^2}{m_e^2}\left(\frac{1}{E_R}-\frac{1}{E_\nu}+\frac{E_R}{4 E_\nu^2}\right) F_{p}^2\left(E_R\right) \times \left |Z_{ion}(E_{er})  \right |^{2}
\end{equation}
where $\mu_{\nu,\rm eff}$ is the effective magnetic moment of solar neutrinos, which comprises a sum of the contributions of each mass eigenstate weighted by the entries of the PMNS matrix \cite{Giunti:2014ixa}

\begin{equation}
\mu_{\nu, \rm eff}^2=\sum_k\left|U_{e k}\right|^2 \sum_j\left|\mu_{j k}-i e_{j k}\right|^2
\end{equation}
where $U$ is the PMNS matrix (which should account for matter effects in the Sun), $\mu_{jk}$ are the magnetic multipoles of the mass eigenstates, and $e_{jk}$ are the electric dipole moments of the mass eigenstates, which is degenerate with the magnetic moment for our recoil rate observable.

Neutrinos are expected to scatter off nuclei also via their anapole moment. In fact, this is the only diagonal moment allowed for Majorana neutrinos. This contribution interferes with the weak current, and the interference can be parametrized via a redefinition of the weak mixing angle in Equation \ref{eq:cenuNs}

\begin{equation}
\sin ^2 \theta_W \rightarrow \sin ^2 \theta_W\left(1-2 m_W^2 a_{\nu, \rm eff}\right)
\end{equation}
where the effective anapole moment is defined similarly as the magnetic moment via

\begin{equation}
a_{\nu, \rm eff}=\sum_k\left|U_{e k}\right| \sum_ja_{jk}
\end{equation}

\begin{figure}[t!]
    \includegraphics[width=0.49\linewidth]{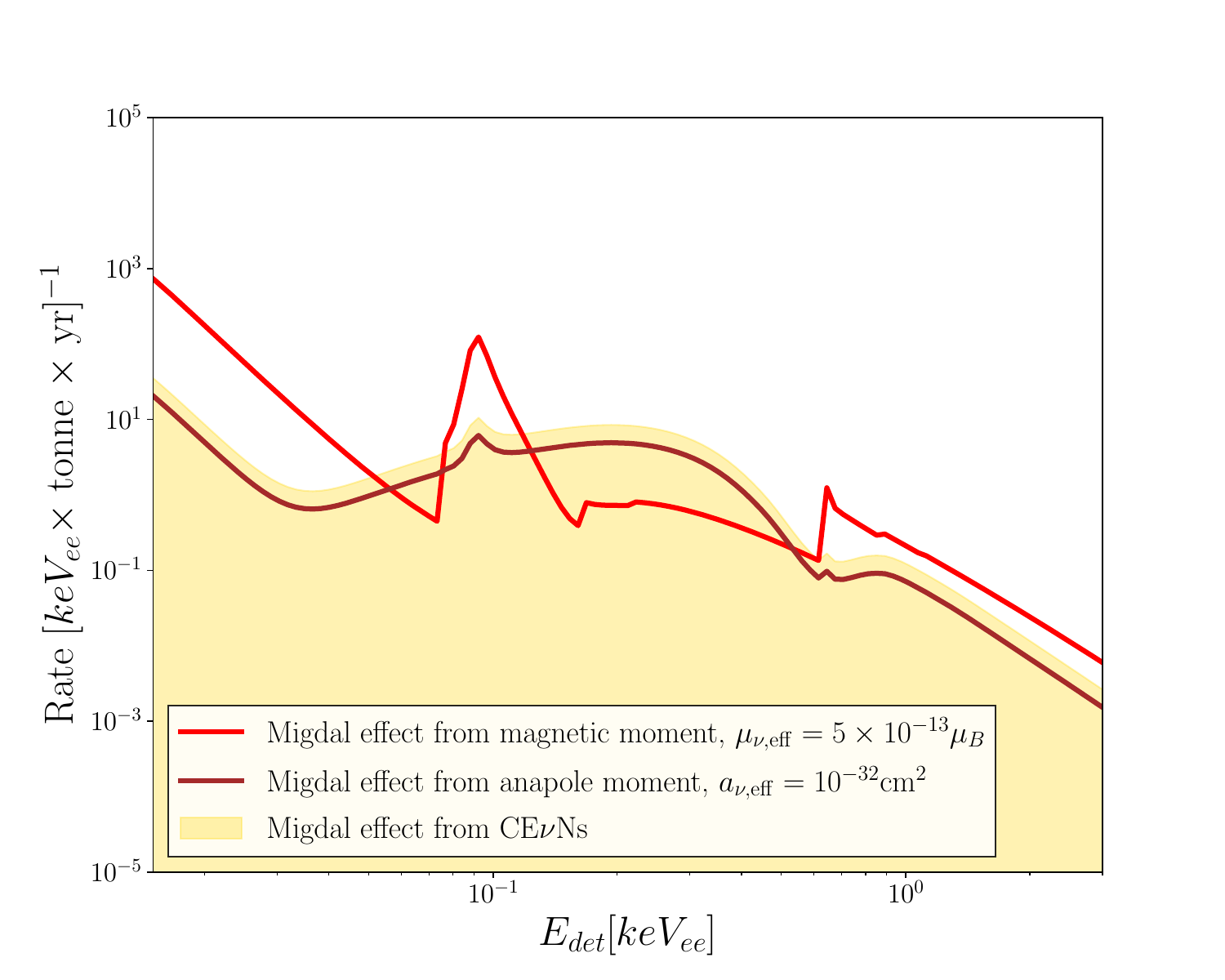}
    \includegraphics[width=0.49\linewidth]{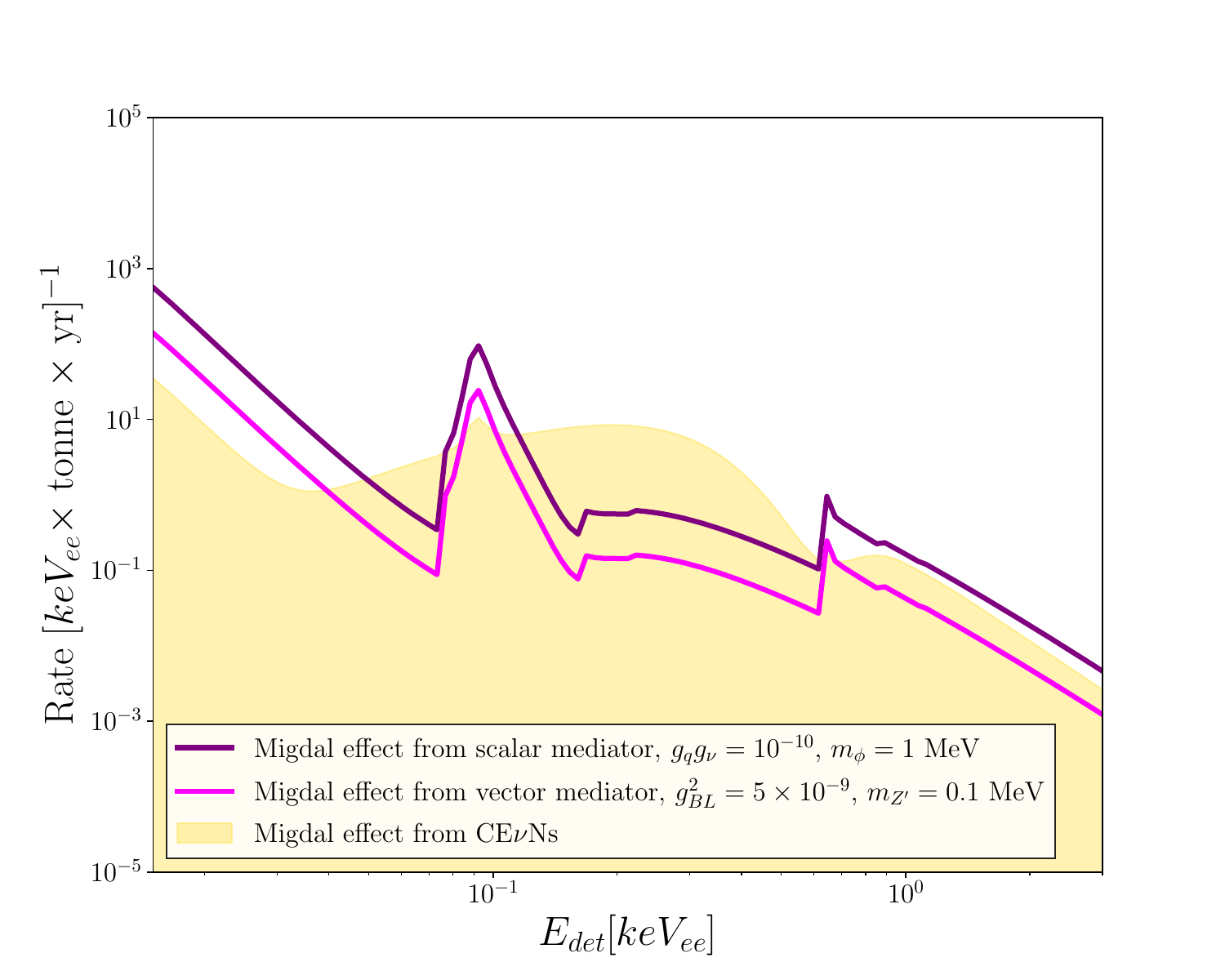}
    \begin{center}
    \includegraphics[width=0.49\linewidth]{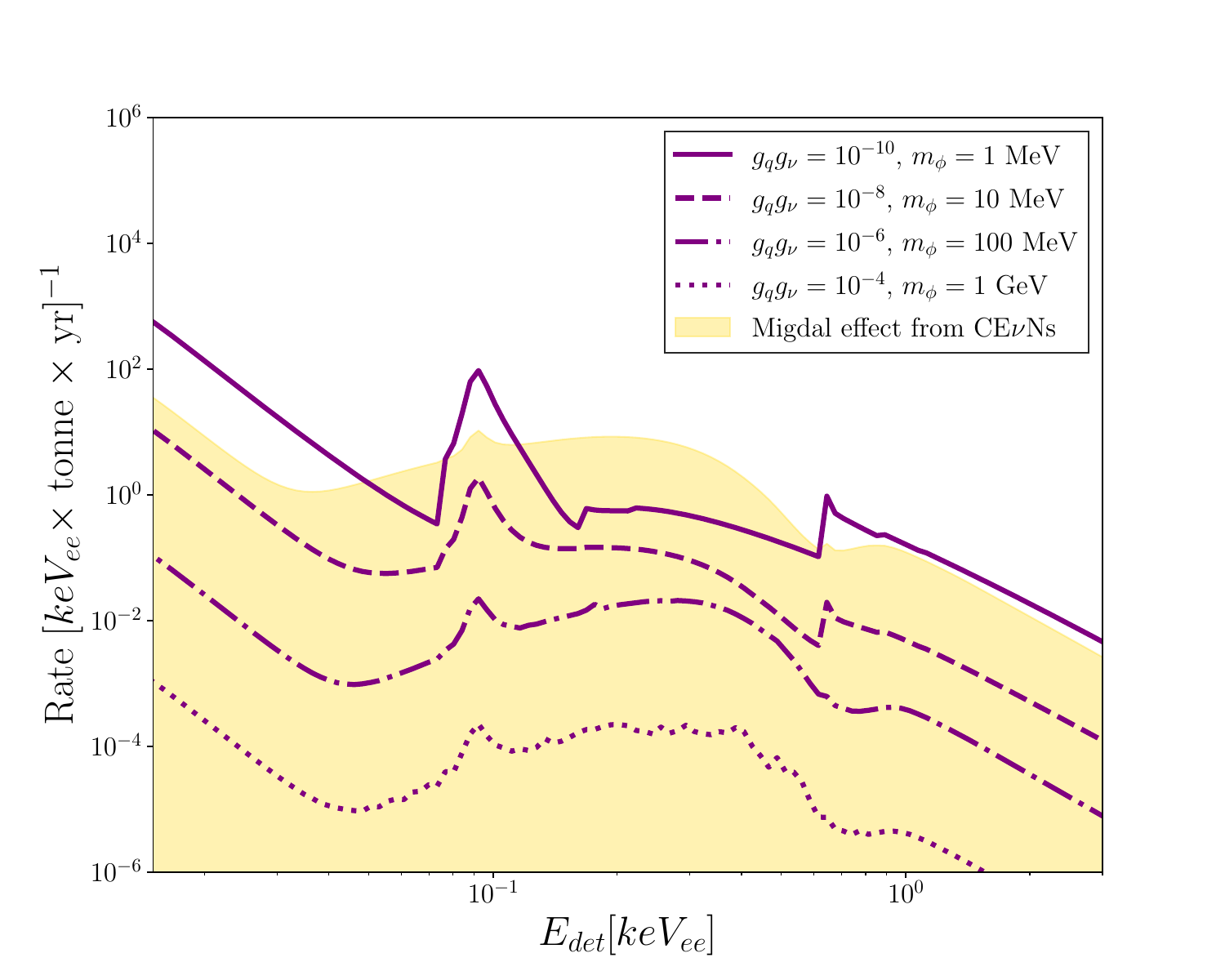}
    \end{center}
\caption{\textit{Upper left plot}: Ionization spectrum of xenon from the Migdal effect due to CE$\nu$NS in the Standard Model (yellow), due to an anapole moment interaction (brown) and due to a magnetic moment interaction (red). For comparison, we show the measured data and background expectation from XENON1T S2-only analysis \cite{Aprile_2019}. \textit{Upper right plot}: Similar plot, but this time confronting the Standard Model expectation with the signal induced by neutrino-nucleus interactions mediated by a new scalar particle (purple) or by a new $B-L$ vector mediator (magenta). \textit{Lower plot:} Ionization spectrum induced by a scalar mediator, for different values of its mass. We illustrate how the peak around 0.1 keV$_{ee}$ is smoothed out for sufficiently heavy mediators, due to the dominance of the more energetic part of the neutrino flux over $pp$ neutrinos.}
\label{fig:Migdal_nonstandard}
\end{figure}

In Figure \ref{fig:Migdal_nonstandard}, we show the electron ionization rate from neutrino-nucleus scattering due to the weak current compared to the magnetic moment contribution and the anapole moment contribution. We show the ionization rates for values which yield rates comparable or larger than the expected contribution from CE$\nu$NS. From this Figure it becomes evident that values of the effective magnetic moment as low as $ \sim 5 \times 10^{-13} \mu_{B}$ and values of the effective anapole moment as low as $ \sim 10^{-32}$cm$^{2}$ could be probed via the Migdal effect in future XENONnT analysis. This value of the magnetic moment might be stronger than current laboratory constraints \cite{XENON:2022ltv,AtzoriCorona:2022jeb}, and the value of the anapole moment is close to the Standard Model values \cite{Cabral-Rosetti:2002zyl}. As expected, the spectral shape of the anapole moment induced ionization rate is the same as the ionization caused by CE$\nu$NS, while the magnetic moment induced ionization rate differs from the Standard Model, due to a dominance of the signal from pp-neutrinos w.r.t the signal from the more energetic neutrino fluxes. This is caused by an enhancement on the neutrino-nucleus scattering cross section via a magnetic moment at low incoming neutrino energies.

The electromagnetic scattering of neutrinos off nuclei is expected to occur in the Standard Model. However, neutrinos may also interact beyond the Standard Model. A simple example consists in neutrinos coupling to quarks via new light scalar or vector mediators. The coherent scattering off nucleus would then also cause an ionization signal due to the Migdal effect. For a scalar mediator there is no interference with the Standard Model contribution, and the Migdal ionization signal from neutrino-nucleus scattering reads \cite{Bertuzzo:2017tuf,Farzan:2018gtr,Miranda:2020tif}:
\begin{equation}
\left(\frac{d \sigma_{\nu N-\nu N^{*} e^{-}}}{d E_R}\right)_{\phi}=\frac{G_F^2 m_A^2}{4 \pi} \frac{g_{\nu \phi} Q_{\phi}^2 E_R}{E_\nu^2\left(2 m_A E_R+m_\phi^2\right)^2} \times \left |Z_{ion}(E_{er})  \right |^{2}
\end{equation}
where the scalar charge is defined as \cite{AristizabalSierra:2019ykk}

\begin{equation}
Q_{\phi}=Z F_p(E_R) \sum_{q=u, d} g_{q\phi} \frac{m_p}{m_q} f_{T_q}^p+N F_n(E_R) \sum_{q=u, d} g_{q\phi} \frac{m_n}{m_q} f_{T_q}^n
\end{equation}
and we take values of the hadronic form factors $f_{T_q}^p$ and $f_{T_q}^n$ from \cite{Hoferichter:2015dsa}. Further, we will assume for simplicity that the scalar $\phi$ couples to quarks $u$ and $d$ with same strength, $g_{u\phi}=g_{d\phi}$.

For the vector mediator case, there is interference with the weak scattering contribution. The interference can be parametrized through a redefinition of the weak vector charge in Equation \ref{eq:cenuNs} \cite{Billard:2018jnl}:

\begin{equation}
Q_V \rightarrow  Q_{Z^{\prime}}=Q_V+\frac{g_{\nu Z^{\prime}}}{\sqrt{2} G_F} \frac{\left(2 g_{uZ^{\prime}}+g_{d Z^{\prime}}\right) Z F_p\left(E_R\right)+\left(g_{uZ^{\prime}}+2 g_{d Z^{\prime}}\right) N F_n\left(E_R\right)}{2 m_A E_R+m_{Z^{\prime}}^2} 
\end{equation}
where again we assume universal quark couplings to the vector mediator, $g_{u Z^{\prime}}=g_{d Z^{\prime}}$. We will focus in a concrete $U(1)_{B-L}$ extension of the Standard Model where the difference of baryon and lepton numbers is gauged. In this case, the coupling of the $Z^{\prime}$ to neutrinos and quarks is related via $g_{q Z^{\prime}}=-g_{\nu Z^{\prime}} / 3$ \cite{Mohapatra,davidson, Billard:2018jnl}.

In Figure \ref{fig:Migdal_nonstandard}, we show the ionization rates induced via the Migdal effect in these models, for values of the couplings and mediator masses currently allowed in parameter space by other experiments. For mediator masses around 1 MeV, strong constraints on couplings of order $g^2 \lesssim 10^{-9}-10^{-10}$ arise from CE$\nu$Ns of reactor neutrinos in nearby detectors \cite{Miranda:2020tif, Lindner:2024eng}, from solar neutrino-electron scattering in Borexino \cite{Agarwalla:2012wf,Bilmis:2015lja,Khan:2019jvr, Coloma:2022umy}, and from solar neutrino-nucleus scatterings in direct detection experiments \cite{Li:2022jfl, Demirci:2023tui, Schwemberger:2023hee, DeRomeri:2024dbv, Amaral:2023tbs}. It can be appreciated in the Figure that for the vector mediator model from a $U(1)_{B-L}$ symmetry, the ionization rate falls below the expected rate in the Standard Model, so a detection of the Migdal effect from CE$\nu$NS with some precision would be needed to set strong constraints in this model. However, for the scalar mediator case, and for currently allowed values of the mediator mass and couplings, the expected scattering rate can be higher than in the Standard Model, which clearly suggests that ionization signals in liquid xenon detectors may be a good probe of these models in the near future. It should be noted, however, that if discrimination of nuclear and electron recoil events at low energies becomes possible, the nuclear recoil signal is expected to provide stronger limits than the Migdal signal, due to the larger total rate. However, the Migdal signal yields a rate with a more peculiar spectral form, which may enhance discrimination from other backgrounds.

As we already mentioned, the spectral shape of the ionization rate due to a neutrino magnetic moment or due to new light mediators differs from the one induced by the weak interactions. This is due to the fact that solar $pp$ neutrinos, with lower energies, contribute predominantly to the ionization signal due to the enhancement in the cross section at low energies in these models. For the Standard Model contribution to the scattering rate, on the other hand, the larger energy neutrino flux from other processes in the Sun, and from the atmospheric and diffuse supernova neutrino background contributes predominantly. The Migdal effect then provides a remarkable feature of new physics in the neutrino sector, which is a peak in the ionization spectrum around 0.1 keV given by the ionization of the electrons in the $n=4$ shell from solar $pp$ neutrinos. This spectral feature differs largely from the scattering rate due to CE$\nu$NS in the Standard Model.

In the lower plot of the Figure we show the ionization rate induced in the scalar mediator model, for different values of the scalar mediator mass. As the mediator mass becomes larger, the contributions to the scattering rate from the neutrino flux on Earth above $\sim 400$ keV energies becomes dominant with respect to the $pp$ neutrinos, and the peak in the spectrum around $\sim 0.1$ keV electron energies is smoothed out substantially.

We believe that near future liquid xenon experiments should be able to resolve such a peak on the spectrum after accounting for energy resolution effects, experimental efficiencies, and background modellings. On the one hand, the main background at such energies in XENON1T is given by events from $\beta$ decays on the cathode wires, which seem to increase linearly at low energies. There is no reason to think that they should peak at 0.1 keV$_{ee}$ with small width. Further, the energy resolution of electron recoil events in XENON1T is smaller than the predicted width of the peak we are discussing, which goes from 0.06 keV$_{ee}$ to 0.12 keV$_{ee}$. The main drawback at this point is the minimal energy threshold and efficiency of these experiments, which currently lies at $\sim 0.2-1$ keV$_{ee}$ for XENON1T and LUX-ZEPLIN \cite{Aprile_2019, LZ:2022lsv}. This energy threshold would need to be reduced by roughly one order of magnitude, while keeping acceptable efficiency levels around 0.1 keV energies. We do not speculate about whether this possibility is feasible in liquid xenon detectors, but we point out that such low-energy thresholds have been achieved in recent years for bolometric detectors \cite{Essig:2022dfa}.

\begin{figure}[t!]\label{fig:Migdal_peak}
    \begin{center}
    \includegraphics[width=0.6\linewidth]{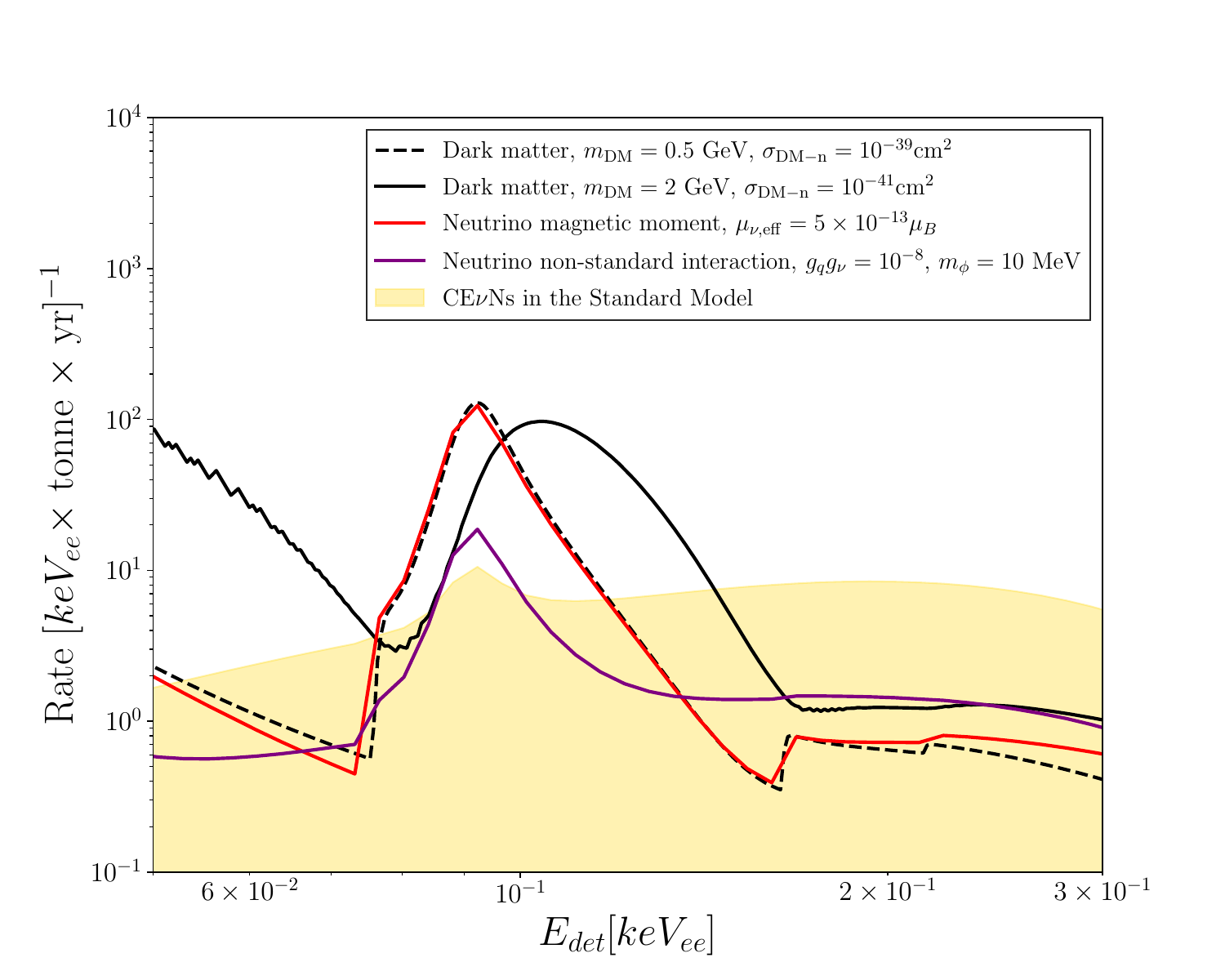}
    \end{center}
\caption{Zoomed ionization spectrum of xenon due to the Migdal effect from CE$\nu$NS around 0.1 keV$_{ee}$. As apparent in the plot, the ionization spectrum in the Standard Model is roughly constant around 0.1 keV$_{ee}$, with a very small peak arising from ionization of electrons in the $n=4$ shell by $pp$ solar neutrinos. However, if neutrinos interact via a magnetic moment interaction, or via new light mediators, the peak can be significantly enhanced w.r.t to the Standard Model. Similarly, for dark matter-nucleus scattering, the Migdal effect induces a peak at those energies, although its width and central value are sensitive to the value of the dark matter mass.}
\label{fig:zoom}
\end{figure}

Of course, the peak may not manifest completely if the cross section of the new physics  model is sufficiently small. Still, yet another contribution to the small peak expected in the Standard Model might be detectable even in this case. A possibility would be to confront the measured spectrum at 0.1 keV$_{ee}$ with the measured spectrum around 1 keV$_{ee}$, where the ionization of electrons in the $n=3$ shell due to new physics may also differ significantly from the Standard Model, as it can be appreciated in our Figures. On the other hand, if a peak is eventually detected, it might be difficult to distinguish whether it arises from neutrino-nucleus scattering or from light dark matter-nucleus scattering. As can be appreciated in Figure \ref{fig:zoom}, dark matter masses below $m_{\rm DM} \lesssim 0.5$ GeV induce a peak at 0.1 keV$_{ee}$ which resembles that induced by a neutrino magnetic moment, for example. If the dark matter mass is heavier, on the other hand, the width and position of the peak can vary significantly. Thus, the resolution of this spectral feature can also be used to constrain the value of the dark matter mass.

\section{Conclusions}\label{sec:conclusions}

The ionization of electrons due to nuclear recoils is a quantum effect predicted in 1939, however, its observation is limited to a few anecdotal phenomena in some nuclear processes. The dubbed "Migdal effect" is expected to cause ionization of electrons on Earth based detectors sensitive to solar neutrinos via their coherent elastic scatterings off nuclei.

In this paper, we have discussed the current sensitivity and prospects for liquid xenon experiments to the ionization signal induced by coherent elastic neutrino-nucleus scattering, further quantifying the irreducible background that it constitutes for light dark matter searches based on the Migdal effect. For dark matter masses in the range from $m_{\rm DM} \sim 0.1-2$ GeV, we find that the neutrino floor is lying from 5 orders of magnitude to less than 4 orders of magnitude below the upper limits derived for the Migdal effect from dark matter by XENON1T.

With current exposures at XENONnT, thresholds as low as those in S2-only data, and assuming background levels as small as in their most recent analysis of electron recoil data, we find that the contribution from the Migdal effect would yield $O(1)$ changes on the detected ionization spectrum from $10^{-1}$ keV$_{ee}$ to 5 $\times 10^{-1}$ keV$_{ee}$, with expected ionization levels due to the Migdal effect that are very close to the background level expectation.

We have also studied the ionization signal induced by neutrino-nucleus scatterings via an effective magnetic and anapole moments, and via new scalar or vector mediators. We find that the anapole moment may exceed the Standard Model contribution to the Migdal ionization rate for values of $a_{\nu,\rm eff} \sim 3 \times 10^{-32}$ cm$^2$, and the magnetic moment contribution may be dominant for values of $\mu_{\nu, \rm eff} \sim 5 \times 10^{-13} \mu_{\rm B}$. For a scalar mediator of mass $m_{\phi} \sim 1-10$ MeV, couplings currently allowed by complementary probes, of $g_{q}g_{\nu} \sim 10^{-11}$, would yield an ionization signal exceeding that in the Standard Model. For a $B-L$ gauged vector mediator, couplings of $g_{BL}^{2} \sim 10^{-8}$ would be needed to overcome the Standard Model expectation.

For the magnetic moment interaction and new mediators with masses below $\sim 10$ MeV, we find a distinct peak in the ionization spectrum of xenon around $0.1$ keV$_{ee}$. This peak arises from the ionizations of electrons in the $n=4$ shell due to solar neutrinos from the $pp$-chain. This peak is substantially depleted in the Standard Model, since the ionization spectrum at these energies is dominated by neutrinos with larger energies than those produced in the $pp$-chain. For these more energetic neutrinos, the ionization spectrum is continuous at $\sim 0.1$ keV$_{ee}$, and the peak is almost completely smoothed out.

The ionization signal induced by the Migdal effect from CE$\nu$NS might be observed in the near future by large exposure dark matter direct detection experiments with low thresholds, becoming an irreducible background for light dark matter searches. Perhaps more importantly, this ionization signal offers a unique way to probe some models of new physics in the neutrino or dark sector, via the appeareance of a prominent peak in the ionization spectrum of xenon around $0.1$ keV$_{ee}$ that is not expected to be present in the Standard Model. We hope that future data from the XENONnT,  LUX-ZEPLIN and PANDA-X experiments will allow to identify the Migdal ionization signal from neutrino-nucleus scatterings, shedding light on the spectrum around 0.1 keV$_{ee}$.

\subsection*{Acknowledgments}

We are grateful to Patrick Huber for useful discussions on the Migdal effect and low-threshold direct detection experiments, and to Duncan Adams and Rouven Essig for discussions on the nuclear uncertainties present in the Migdal effect. We are also grateful to Garv Chauhan, Rijeesh Keloth, Mar Císcar and Camillo Mariani for useful discussions. We also thank Miguel Escudero, Alejandro Ibarra and Gaurav Tomar for useful feedback at the early stages of this project. The work of GH is supported by the U.S. Department of Energy Office of Science under award number DE-SC0020262, by the Collaborative Research Center SFB1258, and by the Deutsche Forschungsgemeinschaft (DFG, German Research Foundation) under Germany's Excellence Strategy - EXC-2094 - 390783311.
\printbibliography

@article{Billard:2013qya,
    author = "Billard, J. and Strigari, L. and Figueroa-Feliciano, E.",
    title = "{Implication of neutrino backgrounds on the reach of next generation dark matter direct detection experiments}",
    eprint = "1307.5458",
    archivePrefix = "arXiv",
    primaryClass = "hep-ph",
    doi = "10.1103/PhysRevD.89.023524",
    journal = "Phys. Rev. D",
    volume = "89",
    number = "2",
    pages = "023524",
    year = "2014"
}

@article{Bernabei:2007jz,
    author = "Bernabei, R. and others",
    title = "{On electromagnetic contributions in WIMP quests}",
    eprint = "0706.1421",
    archivePrefix = "arXiv",
    primaryClass = "astro-ph",
    doi = "10.1142/S0217751X07037093",
    journal = "Int. J. Mod. Phys. A",
    volume = "22",
    pages = "3155--3168",
    year = "2007"
}

@article{Dolan:2017xbu,
    author = "Dolan, Matthew J. and Kahlhoefer, Felix and McCabe, Christopher",
    title = "{Directly detecting sub-GeV dark matter with electrons from nuclear scattering}",
    eprint = "1711.09906",
    archivePrefix = "arXiv",
    primaryClass = "hep-ph",
    reportNumber = "KCL-PH-TH/2017-54, TTK-17-43, KCL-PH-TH-2017-54",
    doi = "10.1103/PhysRevLett.121.101801",
    journal = "Phys. Rev. Lett.",
    volume = "121",
    number = "10",
    pages = "101801",
    year = "2018"
}

@article{Essig:2019xkx,
    author = "Essig, Rouven and Pradler, Josef and Sholapurkar, Mukul and Yu, Tien-Tien",
    title = "{Relation between the Migdal Effect and Dark Matter-Electron Scattering in Isolated Atoms and Semiconductors}",
    eprint = "1908.10881",
    archivePrefix = "arXiv",
    primaryClass = "hep-ph",
    reportNumber = "YITP-19-23",
    doi = "10.1103/PhysRevLett.124.021801",
    journal = "Phys. Rev. Lett.",
    volume = "124",
    number = "2",
    pages = "021801",
    year = "2020"
}

@article{Baxter:2019pnz,
    author = "Baxter, Daniel and Kahn, Yonatan and Krnjaic, Gordan",
    title = "{Electron Ionization via Dark Matter-Electron Scattering and the Migdal Effect}",
    eprint = "1908.00012",
    archivePrefix = "arXiv",
    primaryClass = "hep-ph",
    reportNumber = "FERMILAB-PUB-19-257-A",
    doi = "10.1103/PhysRevD.101.076014",
    journal = "Phys. Rev. D",
    volume = "101",
    number = "7",
    pages = "076014",
    year = "2020"
}

@article{Knapen:2020aky,
    author = "Knapen, Simon and Kozaczuk, Jonathan and Lin, Tongyan",
    title = "{Migdal Effect in Semiconductors}",
    eprint = "2011.09496",
    archivePrefix = "arXiv",
    primaryClass = "hep-ph",
    doi = "10.1103/PhysRevLett.127.081805",
    journal = "Phys. Rev. Lett.",
    volume = "127",
    number = "8",
    pages = "081805",
    year = "2021"
}

@article{Liu:2020pat,
    author = "Liu, C. P. and Wu, Chih-Pan and Chi, Hsin-Chang and Chen, Jiunn-Wei",
    title = "{Model-independent determination of the Migdal effect via photoabsorption}",
    eprint = "2007.10965",
    archivePrefix = "arXiv",
    primaryClass = "hep-ph",
    doi = "10.1103/PhysRevD.102.121303",
    journal = "Phys. Rev. D",
    volume = "102",
    number = "12",
    pages = "121303",
    year = "2020"
}

@article{Tomar:2022ofh,
    author = "Tomar, Gaurav and Kang, Sunghyun and Scopel, Stefano",
    title = "{Low-mass extension of direct detection bounds on WIMP-quark and WIMP-gluon effective interactions using the Migdal effect}",
    eprint = "2210.00199",
    archivePrefix = "arXiv",
    primaryClass = "hep-ph",
    reportNumber = "CQUeST-2022-0698",
    doi = "10.1016/j.astropartphys.2023.102851",
    journal = "Astropart. Phys.",
    volume = "150",
    pages = "102851",
    year = "2023"
}

@article{Berghaus:2022pbu,
    author = "Berghaus, Kim V. and Esposito, Angelo and Essig, Rouven and Sholapurkar, Mukul",
    title = "{The Migdal effect in semiconductors for dark matter with masses below \ensuremath{\sim} 100 MeV}",
    eprint = "2210.06490",
    archivePrefix = "arXiv",
    primaryClass = "hep-ph",
    doi = "10.1007/JHEP01(2023)023",
    journal = "JHEP",
    volume = "01",
    pages = "023",
    year = "2023"
}

@article{Cox:2022ekg,
    author = "Cox, Peter and Dolan, Matthew J. and McCabe, Christopher and Quiney, Harry M.",
    title = "{Precise predictions and new insights for atomic ionization from the Migdal effect}",
    eprint = "2208.12222",
    archivePrefix = "arXiv",
    primaryClass = "hep-ph",
    reportNumber = "KCL-2022-30",
    doi = "10.1103/PhysRevD.107.035032",
    journal = "Phys. Rev. D",
    volume = "107",
    number = "3",
    pages = "035032",
    year = "2023"
}

@article{XENON:2019zpr,
    author = "Aprile, E. and others",
    collaboration = "XENON",
    title = "{Search for Light Dark Matter Interactions Enhanced by the Migdal Effect or Bremsstrahlung in XENON1T}",
    eprint = "1907.12771",
    archivePrefix = "arXiv",
    primaryClass = "hep-ex",
    doi = "10.1103/PhysRevLett.123.241803",
    journal = "Phys. Rev. Lett.",
    volume = "123",
    number = "24",
    pages = "241803",
    year = "2019"
}

@article{Harnik:2012ni,
    author = "Harnik, Roni and Kopp, Joachim and Machado, Pedro A. N.",
    title = "{Exploring nu Signals in Dark Matter Detectors}",
    eprint = "1202.6073",
    archivePrefix = "arXiv",
    primaryClass = "hep-ph",
    reportNumber = "FERMILAB-PUB-12-048-T",
    doi = "10.1088/1475-7516/2012/07/026",
    journal = "JCAP",
    volume = "07",
    pages = "026",
    year = "2012"
}

@article{Billard:2014yka,
    author = "Billard, J. and Strigari, L. E. and Figueroa-Feliciano, E.",
    title = "{Solar neutrino physics with low-threshold dark matter detectors}",
    eprint = "1409.0050",
    archivePrefix = "arXiv",
    primaryClass = "astro-ph.CO",
    doi = "10.1103/PhysRevD.91.095023",
    journal = "Phys. Rev. D",
    volume = "91",
    number = "9",
    pages = "095023",
    year = "2015"
}

@article{Ibe:2017yqa,
    author = "Ibe, Masahiro and Nakano, Wakutaka and Shoji, Yutaro and Suzuki, Kazumine",
    title = "{Migdal Effect in Dark Matter Direct Detection Experiments}",
    eprint = "1707.07258",
    archivePrefix = "arXiv",
    primaryClass = "hep-ph",
    reportNumber = "IPMU17-0100",
    doi = "10.1007/JHEP03(2018)194",
    journal = "JHEP",
    volume = "03",
    pages = "194",
    year = "2018"
}

@article{Bell:2019egg,
    author = "Bell, Nicole F. and Dent, James B. and Newstead, Jayden L. and Sabharwal, Subir and Weiler, Thomas J.",
    title = "{Migdal effect and photon bremsstrahlung in effective field theories of dark matter direct detection and coherent elastic neutrino-nucleus scattering}",
    eprint = "1905.00046",
    archivePrefix = "arXiv",
    primaryClass = "hep-ph",
    doi = "10.1103/PhysRevD.101.015012",
    journal = "Phys. Rev. D",
    volume = "101",
    number = "1",
    pages = "015012",
    year = "2020"
}

@article{Migdal1977QualitativeMI,
  title={Qualitative Methods in Quantum Theory},
  author={A. B. Migdal},
  journal={Frontiers in Physiology},
  year={1977},
  volume={48},
  pages={1-437}
}

@article{Aprile_2019,
	doi = {10.1103/physrevlett.123.251801},
  
	url = {https://doi.org/10.1103%2Fphysrevlett.123.251801},
  
	year = 2019,
	month = {dec},
  
	publisher = {American Physical Society ({APS})},
  
	volume = {123},
  
	number = {25},
  
	author = {E. Aprile and J. Aalbers and F. Agostini and M. Alfonsi and L. Althueser and F.{\hspace{0.167em}
}D. Amaro and V.{\hspace{0.167em}}C. Antochi and E. Angelino and F. Arneodo and D. Barge and L. Baudis and B. Bauermeister and L. Bellagamba and M.{\hspace{0.167em}}L. Benabderrahmane and T. Berger and P.{\hspace{0.167em}}A. Breur and A. Brown and E. Brown and S. Bruenner and G. Bruno and R. Budnik and C. Capelli and J.{\hspace{0.167em}}M.{\hspace{0.167em}}R. Cardoso and D. Cichon and D. Coderre and A.{\hspace{0.167em}}P. Colijn and J. Conrad and J.{\hspace{0.167em}}P. Cussonneau and M.{\hspace{0.167em}}P. Decowski and P. de Perio and A. Depoian and P. Di Gangi and A. Di Giovanni and S. Diglio and A. Elykov and G. Eurin and J. Fei and A.{\hspace{0.167em}}D. Ferella and A. Fieguth and W. Fulgione and P. Gaemers and A. Gallo Rosso and M. Galloway and F. Gao and M. Garbini and L. Grandi and Z. Greene and C. Hasterok and C. Hils and E. Hogenbirk and J. Howlett and M. Iacovacci and R. Itay and F. Joerg and S. Kazama and A. Kish and M. Kobayashi and G. Koltman and A. Kopec and H. Landsman and R.{\hspace{0.167em}}F. Lang and L. Levinson and Q. Lin and S. Lindemann and M. Lindner and F. Lombardi and J.{\hspace{0.167em}}A.{\hspace{0.167em}}M. Lopes and E. L{\'{o}}pez Fune and C. Macolino and J. Mahlstedt and A. Manfredini and F. Marignetti and T. Marrod{\'{a}}n Undagoitia and J. Masbou and S. Mastroianni and M. Messina and K. Micheneau and K. Miller and A. Molinario and K. Mor{\aa} and Y. Mosbacher and M. Murra and J. Naganoma and K. Ni and U. Oberlack and K. Odgers and J. Palacio and B. Pelssers and R. Peres and J. Pienaar and V. Pizzella and G. Plante and R. Podviianiuk and J. Qin and H. Qiu and D. Ram{\'{\i}}rez Garc{\'{\i}}a and S. Reichard and B. Riedel and A. Rocchetti and N. Rupp and J.{\hspace{0.167em}}M.{\hspace{0.167em}}F. dos Santos and G. Sartorelli and N. {\v{S}}ar{\v{c}}evi{\'{c}} and M. Scheibelhut and S. Schindler and J. Schreiner and D. Schulte and M. Schumann and L. Scotto Lavina and M. Selvi and P. Shagin and E. Shockley and M. Silva and H. Simgen and C. Therreau and D. Thers and F. Toschi and G. Trinchero and C. Tunnell and N. Upole and M. Vargas and G. Volta and O. Wack and H. Wang and Y. Wei and C. Weinheimer and D. Wenz and C. Wittweg and J. Wulf and J. Ye and Y. Zhang and T. Zhu and J.{\hspace{0.167em}}P. Zopounidis and},
  
	title = {Light Dark Matter Search with Ionization Signals in {XENON}1T},
  
	journal = {Physical Review Letters}
}

@article{Aprile:2022vux,
    author = "Aprile, E. and others",
    title = "{Search for New Physics in Electronic Recoil Data from XENONnT}",
    eprint = "2207.11330",
    archivePrefix = "arXiv",
    primaryClass = "hep-ex",
    month = "7",
    year = "2022"
}

@article{XENON:2022mpc,
    author = "Aprile, E. and others",
    collaboration = "XENON",
    title = "{Search for New Physics in Electronic Recoil Data from XENONnT}",
    eprint = "2207.11330",
    archivePrefix = "arXiv",
    primaryClass = "hep-ex",
    month = "7",
    year = "2022"
}

@article{Freedman:1973yd,
    author = "Freedman, Daniel Z.",
    title = "{Coherent Neutrino Nucleus Scattering as a Probe of the Weak Neutral Current}",
    reportNumber = "NAL-PUB-73-76-THY, FERMILAB-PUB-73-076-T",
    doi = "10.1103/PhysRevD.9.1389",
    journal = "Phys. Rev. D",
    volume = "9",
    pages = "1389--1392",
    year = "1974"
}

@article{LUX:2018akb,
    author = "Akerib, D. S. and others",
    collaboration = "LUX",
    title = "{Results of a Search for Sub-GeV Dark Matter Using 2013 LUX Data}",
    eprint = "1811.11241",
    archivePrefix = "arXiv",
    primaryClass = "astro-ph.CO",
    doi = "10.1103/PhysRevLett.122.131301",
    journal = "Phys. Rev. Lett.",
    volume = "122",
    number = "13",
    pages = "131301",
    year = "2019"
}

@article{CDEX:2019hzn,
    author = "Liu, Z. Z. and others",
    collaboration = "CDEX",
    title = "{Constraints on Spin-Independent Nucleus Scattering with sub-GeV Weakly Interacting Massive Particle Dark Matter from the CDEX-1B Experiment at the China Jinping Underground Laboratory}",
    eprint = "1905.00354",
    archivePrefix = "arXiv",
    primaryClass = "hep-ex",
    doi = "10.1103/PhysRevLett.123.161301",
    journal = "Phys. Rev. Lett.",
    volume = "123",
    number = "16",
    pages = "161301",
    year = "2019"
}

@article{Armengaud_2022,
   title={Search for sub-GeV dark matter via the Migdal effect with an EDELWEISS germanium detector with NbSi transition-edge sensors},
   volume={106},
   ISSN={2470-0029},
   url={http://dx.doi.org/10.1103/PhysRevD.106.062004},
   DOI={10.1103/physrevd.106.062004},
   number={6},
   journal={Physical Review D},
   publisher={American Physical Society (APS)},
   author={Armengaud, E. and Arnaud, Q. and Augier, C. and Benoît, A. and Bergé, L. and Billard, J. and Broniatowski, A. and Camus, P. and Cazes, A. and Chapellier, M. and Charlieux, F. and De Jésus, M. and Dumoulin, L. and Eitel, K. and Filippini, J.-B. and Filosofov, D. and Gascon, J. and Giuliani, A. and Gros, M. and Guy, E. and Jin, Y. and Juillard, A. and Kleifges, M. and Lattaud, H. and Marnieros, S. and Misiak, D. and Navick, X.-F. and Nones, C. and Olivieri, E. and Oriol, C. and Pari, P. and Paul, B. and Poda, D. and Rozov, S. and Salagnac, T. and Sanglard, V. and Vagneron, L. and Yakushev, E. and Zolotarova, A. and Kavanagh, B. J.},
   year={2022},
   month=sep }

@article{Vergados:2005dpd,
    author = "Vergados, J. D. and Ejiri, H.",
    title = "{The role of ionization electrons in direct neutralino detection}",
    eprint = "hep-ph/0401151",
    archivePrefix = "arXiv",
    doi = "10.1016/j.physletb.2004.11.085",
    journal = "Phys. Lett. B",
    volume = "606",
    pages = "313--322",
    year = "2005"
}

@article{Moustakidis:2005gx,
    author = "Moustakidis, Ch. C. and Vergados, J. D. and Ejiri, H.",
    title = "{Direct dark matter detection by observing electrons produced in neutralino-nucleus collisions}",
    eprint = "hep-ph/0507123",
    archivePrefix = "arXiv",
    doi = "10.1016/j.nuclphysb.2005.08.033",
    journal = "Nucl. Phys. B",
    volume = "727",
    pages = "406--420",
    year = "2005"
}

@article{XENON:2023cxc,
    author = "Aprile, E. and others",
    collaboration = "XENON",
    title = "{First Dark Matter Search with Nuclear Recoils from the XENONnT Experiment}",
    eprint = "2303.14729",
    archivePrefix = "arXiv",
    primaryClass = "hep-ex",
    doi = "10.1103/PhysRevLett.131.041003",
    journal = "Phys. Rev. Lett.",
    volume = "131",
    number = "4",
    pages = "041003",
    year = "2023"
}

@article{LZ:2022lsv,
    author = "Aalbers, J. and others",
    collaboration = "LZ",
    title = "{First Dark Matter Search Results from the LUX-ZEPLIN (LZ) Experiment}",
    eprint = "2207.03764",
    archivePrefix = "arXiv",
    primaryClass = "hep-ex",
    doi = "10.1103/PhysRevLett.131.041002",
    journal = "Phys. Rev. Lett.",
    volume = "131",
    number = "4",
    pages = "041002",
    year = "2023"
}

@article{Miranda:2020tif,
    author = "Miranda, O. G. and Papoulias, D. K. and Sanchez Garcia, G. and Sanders, O. and T\'ortola, M. and Valle, J. W. F.",
    title = "{Implications of the first detection of coherent elastic neutrino-nucleus scattering (CEvNS) with Liquid Argon}",
    eprint = "2003.12050",
    archivePrefix = "arXiv",
    primaryClass = "hep-ph",
    reportNumber = "IFIC/20-XXX",
    doi = "10.1007/JHEP05(2020)130",
    journal = "JHEP",
    volume = "05",
    pages = "130",
    year = "2020",
    note = "[Erratum: JHEP 01, 067 (2021)]"
}

@article{AristizabalSierra:2019ykk,
    author = "Aristizabal Sierra, D. and Dutta, Bhaskar and Liao, Shu and Strigari, Louis E.",
    title = "{Coherent elastic neutrino-nucleus scattering in multi-ton scale dark matter experiments: Classification of vector and scalar interactions new physics signals}",
    eprint = "1910.12437",
    archivePrefix = "arXiv",
    primaryClass = "hep-ph",
    doi = "10.1007/JHEP12(2019)124",
    journal = "JHEP",
    volume = "12",
    pages = "124",
    year = "2019"
}

@article{Hoferichter:2015dsa,
    author = "Hoferichter, Martin and Ruiz de Elvira, J. and Kubis, Bastian and Mei\ss{}ner, Ulf-G.",
    title = "{High-Precision Determination of the Pion-Nucleon \ensuremath{\sigma} Term from Roy-Steiner Equations}",
    eprint = "1506.04142",
    archivePrefix = "arXiv",
    primaryClass = "hep-ph",
    reportNumber = "INT-PUB-15-026",
    doi = "10.1103/PhysRevLett.115.092301",
    journal = "Phys. Rev. Lett.",
    volume = "115",
    pages = "092301",
    year = "2015"
}

@article{Billard:2018jnl,
    author = "Billard, Julien and Johnston, Joseph and Kavanagh, Bradley J.",
    title = "{Prospects for exploring New Physics in Coherent Elastic Neutrino-Nucleus Scattering}",
    eprint = "1805.01798",
    archivePrefix = "arXiv",
    primaryClass = "hep-ph",
    doi = "10.1088/1475-7516/2018/11/016",
    journal = "JCAP",
    volume = "11",
    pages = "016",
    year = "2018"
}

@article{Farzan:2018gtr,
    author = "Farzan, Yasaman and Lindner, Manfred and Rodejohann, Werner and Xu, Xun-Jie",
    title = "{Probing neutrino coupling to a light scalar with coherent neutrino scattering}",
    eprint = "1802.05171",
    archivePrefix = "arXiv",
    primaryClass = "hep-ph",
    doi = "10.1007/JHEP05(2018)066",
    journal = "JHEP",
    volume = "05",
    pages = "066",
    year = "2018"
}

@article{Bertuzzo:2017tuf,
    author = "Bertuzzo, Enrico and Deppisch, Frank F. and Kulkarni, Suchita and Perez Gonzalez, Yuber F. and Zukanovich Funchal, Renata",
    title = "{Dark Matter and Exotic Neutrino Interactions in Direct Detection Searches}",
    eprint = "1701.07443",
    archivePrefix = "arXiv",
    primaryClass = "hep-ph",
    doi = "10.1007/JHEP04(2017)073",
    journal = "JHEP",
    volume = "04",
    pages = "073",
    year = "2017"
}

@article{Giunti:2014ixa,
    author = "Giunti, Carlo and Studenikin, Alexander",
    title = "{Neutrino electromagnetic interactions: a window to new physics}",
    eprint = "1403.6344",
    archivePrefix = "arXiv",
    primaryClass = "hep-ph",
    doi = "10.1103/RevModPhys.87.531",
    journal = "Rev. Mod. Phys.",
    volume = "87",
    pages = "531",
    year = "2015"
}

@article{Green:2011bv,
    author = "Green, Anne M.",
    title = "{Astrophysical uncertainties on direct detection experiments}",
    eprint = "1112.0524",
    archivePrefix = "arXiv",
    primaryClass = "astro-ph.CO",
    doi = "10.1142/S0217732312300042",
    journal = "Mod. Phys. Lett. A",
    volume = "27",
    pages = "1230004",
    year = "2012"
}

@article{Coloma:2022umy,
    author = "Coloma, Pilar and Coloma, Pilar and Gonzalez-Garcia, M. C. and Gonzalez-Garcia, M. C. and Maltoni, Michele and Maltoni, Michele and Pinheiro, Jo\~ao Paulo and Pinheiro, Jo\~ao Paulo and Urrea, Salvador and Urrea, Salvador",
    title = "{Constraining new physics with Borexino Phase-II spectral data}",
    eprint = "2204.03011",
    archivePrefix = "arXiv",
    primaryClass = "hep-ph",
    reportNumber = "IFT-UAM/CSIC-22-14, IFIC/22-15, FTUV-22-0404.7998, YITP-SB-2022-05",
    doi = "10.1007/JHEP07(2022)138",
    journal = "JHEP",
    volume = "07",
    pages = "138",
    year = "2022",
    note = "[Erratum: JHEP 11, 138 (2022)]"
}

@article{Deason_2019,
   title={The local high-velocity tail and the Galactic escape speed},
   volume={485},
   ISSN={1365-2966},
   url={http://dx.doi.org/10.1093/mnras/stz623},
   DOI={10.1093/mnras/stz623},
   number={3},
   journal={Monthly Notices of the Royal Astronomical Society},
   publisher={Oxford University Press (OUP)},
   author={Deason, Alis J and Fattahi, Azadeh and Belokurov, Vasily and Evans, N Wyn and Grand, Robert J J and Marinacci, Federico and Pakmor, Rüdiger},
   year={2019},
   month=mar, pages={3514–3526} }

@article{Herrera:2021puj,
    author = "Herrera, Gonzalo and Ibarra, Alejandro",
    title = "{Direct detection of non-galactic light dark matter}",
    eprint = "2104.04445",
    archivePrefix = "arXiv",
    primaryClass = "hep-ph",
    reportNumber = "TUM-HEP 1320/21",
    doi = "10.1016/j.physletb.2021.136551",
    journal = "Phys. Lett. B",
    volume = "820",
    pages = "136551",
    year = "2021"
}

@article{Bell:2021ihi,
    author = "Bell, Nicole F. and Dent, James B. and Lang, Rafael F. and Newstead, Jayden L. and Ritter, Alexander C.",
    title = "{Observing the Migdal effect from nuclear recoils of neutral particles with liquid xenon and argon detectors}",
    eprint = "2112.08514",
    archivePrefix = "arXiv",
    primaryClass = "hep-ph",
    doi = "10.1103/PhysRevD.105.096015",
    journal = "Phys. Rev. D",
    volume = "105",
    number = "9",
    pages = "096015",
    year = "2022"
}

@article{Li:2023xkf,
    author = "Li, Yu-Feng and Xia, Shuo-yu",
    title = "{Migdal effect of Phonon-mediated neutrino nucleus scattering in semiconductor detectors}",
    eprint = "2310.05704",
    archivePrefix = "arXiv",
    primaryClass = "hep-ph",
    month = "10",
    year = "2023"
}

@article{Xu:2023wev,
    author = "Xu, Jingke and others",
    title = "{Search for the Migdal effect in liquid xenon with keV-level nuclear recoils}",
    eprint = "2307.12952",
    archivePrefix = "arXiv",
    primaryClass = "hep-ex",
    reportNumber = "LLNL-JRNL-850170",
    month = "7",
    year = "2023"
}

@article{Bell:2023uvf,
    author = "Bell, Nicole F. and Cox, Peter and Dolan, Matthew J. and Newstead, Jayden L. and Ritter, Alexander C.",
    title = "{Exploring light dark matter with the Migdal effect in hydrogen-doped liquid xenon}",
    eprint = "2305.04690",
    archivePrefix = "arXiv",
    primaryClass = "hep-ph",
    month = "5",
    year = "2023"
}

@article{Adams:2022zvg,
    author = "Adams, Duncan and Baxter, Daniel and Day, Hannah and Essig, Rouven and Kahn, Yonatan",
    title = "{Measuring the Migdal effect in semiconductors for dark matter detection}",
    eprint = "2210.04917",
    archivePrefix = "arXiv",
    primaryClass = "hep-ph",
    reportNumber = "FERMILAB-PUB-22-705-PPD-QIS-T",
    doi = "10.1103/PhysRevD.107.L041303",
    journal = "Phys. Rev. D",
    volume = "107",
    number = "4",
    pages = "L041303",
    year = "2023"
}

@article{Gu:2023pfg,
    author = "Gu, Yuchao and Tang, Jie and Wu, Lei and Zhu, Bin",
    title = "{Probing Light DM through Migdal Effect with Spherical Proportional Counter}",
    eprint = "2309.09740",
    archivePrefix = "arXiv",
    primaryClass = "hep-ph",
    month = "9",
    year = "2023"
}

@article{Blanco:2022pkt,
    author = "Blanco, Carlos and Harris, Ian and Kahn, Yonatan and Lillard, Benjamin and P\'erez-R\'\i{}os, Jes\'us",
    title = "{Molecular Migdal effect}",
    eprint = "2208.09002",
    archivePrefix = "arXiv",
    primaryClass = "hep-ph",
    doi = "10.1103/PhysRevD.106.115015",
    journal = "Phys. Rev. D",
    volume = "106",
    number = "11",
    pages = "115015",
    year = "2022"
}

@article{Wang:2021oha,
    author = "Wang, Wenyu and Wu, Ke-Yun and Wu, Lei and Zhu, Bin",
    title = "{Direct detection of spin-dependent sub-GeV dark matter via Migdal effect}",
    eprint = "2112.06492",
    archivePrefix = "arXiv",
    primaryClass = "hep-ph",
    doi = "10.1016/j.nuclphysb.2022.115907",
    journal = "Nucl. Phys. B",
    volume = "983",
    pages = "115907",
    year = "2022"
}

@ARTICLE{compton_anomaly,
       author = {{Karlsson}, E.~B. and {Hartmann}, O. and {Chatzidimitriou-Dreismann}, C.~A. and {Abdul-Redah}, T.},
        title = "{The hydrogen anomaly in neutron Compton scattering: new experiments and a quantitative theoretical explanation}",
      journal = {Measurement Science and Technology},
         year = 2016,
        month = aug,
       volume = {27},
       number = {8},
          eid = {085501},
        pages = {085501},
          doi = {10.1088/0957-0233/27/8/085501},
       adsurl = {https://ui.adsabs.harvard.edu/abs/2016MeScT..27h5501K}
}

@article{PhysRevLett.108.243201,
  title = {First Measurement of Pure Electron Shakeoff in the $\ensuremath{\beta}$ Decay of Trapped $^{6}\mathrm{He}^{+}$ Ions},
  author = {Couratin, C. and Velten, Ph. and Fl\'echard, X. and Li\'enard, E. and Ban, G. and Cassimi, A. and Delahaye, P. and Durand, D. and Hennecart, D. and Mauger, F. and M\'ery, A. and Naviliat-Cuncic, O. and Patyk, Z. and Rodr\'{\i}guez, D. and Siegie\ifmmode \acute{n}\else \'{n}\fi{}-Iwaniuk, K. and Thomas, J-C.},
  journal = {Phys. Rev. Lett.},
  volume = {108},
  issue = {24},
  pages = {243201},
  numpages = {4},
  year = {2012},
  month = {Jun},
  publisher = {American Physical Society},
  doi = {10.1103/PhysRevLett.108.243201},
  url = {https://link.aps.org/doi/10.1103/PhysRevLett.108.243201}
}

@article{Nakamura:2020kex,
    author = "Nakamura, Kiseki D. and Miuchi, Kentaro and Kazama, Shingo and Shoji, Yutaro and Ibe, Masahiro and Nakano, Wakutaka",
    title = "{Detection capability of the Migdal effect for argon and xenon nuclei with position-sensitive gaseous detectors}",
    eprint = "2009.05939",
    archivePrefix = "arXiv",
    primaryClass = "physics.ins-det",
    doi = "10.1093/ptep/ptaa162",
    journal = "PTEP",
    volume = "2021",
    number = "1",
    pages = "013C01",
    year = "2021"
}

@article{Araujo:2022wjh,
    author = "Ara\'ujo, H. M. and others",
    title = "{The MIGDAL experiment: Measuring a rare atomic process to aid the search for dark matter}",
    eprint = "2207.08284",
    archivePrefix = "arXiv",
    primaryClass = "hep-ex",
    doi = "10.1016/j.astropartphys.2023.102853",
    journal = "Astropart. Phys.",
    volume = "151",
    pages = "102853",
    year = "2023"
}

@article{OHare:2021utq,
    author = "O'Hare, Ciaran A. J.",
    title = "{New Definition of the Neutrino Floor for Direct Dark Matter Searches}",
    eprint = "2109.03116",
    archivePrefix = "arXiv",
    primaryClass = "hep-ph",
    doi = "10.1103/PhysRevLett.127.251802",
    journal = "Phys. Rev. Lett.",
    volume = "127",
    number = "25",
    pages = "251802",
    year = "2021"
}

@article{Dent:2016wcr,
    author = "Dent, James B. and Dutta, Bhaskar and Liao, Shu and Newstead, Jayden L. and Strigari, Louis E. and Walker, Joel W.",
    title = "{Probing light mediators at ultralow threshold energies with coherent elastic neutrino-nucleus scattering}",
    eprint = "1612.06350",
    archivePrefix = "arXiv",
    primaryClass = "hep-ph",
    doi = "10.1103/PhysRevD.96.095007",
    journal = "Phys. Rev. D",
    volume = "96",
    number = "9",
    pages = "095007",
    year = "2017"
}

@article{Dutta:2017nht,
    author = "Dutta, Bhaskar and Liao, Shu and Strigari, Louis E. and Walker, Joel W.",
    title = "{Non-standard interactions of solar neutrinos in dark matter experiments}",
    eprint = "1705.00661",
    archivePrefix = "arXiv",
    primaryClass = "hep-ph",
    reportNumber = "MI-TH-1724",
    doi = "10.1016/j.physletb.2017.08.031",
    journal = "Phys. Lett. B",
    volume = "773",
    pages = "242--246",
    year = "2017"
}

@article{AristizabalSierra:2017joc,
    author = "Aristizabal Sierra, D. and Rojas, N. and Tytgat, M. H. G.",
    title = "{Neutrino non-standard interactions and dark matter searches with multi-ton scale detectors}",
    eprint = "1712.09667",
    archivePrefix = "arXiv",
    primaryClass = "hep-ph",
    doi = "10.1007/JHEP03(2018)197",
    journal = "JHEP",
    volume = "03",
    pages = "197",
    year = "2018"
}

@article{Cerdeno:2016sfi,
    author = "Cerde\~no, David G. and Fairbairn, Malcolm and Jubb, Thomas and Machado, Pedro A. N. and Vincent, Aaron C. and B\oe{}hm, C\'eline",
    title = "{Physics from solar neutrinos in dark matter direct detection experiments}",
    eprint = "1604.01025",
    archivePrefix = "arXiv",
    primaryClass = "hep-ph",
    reportNumber = "IFT-UAM-CSIC-16-031, FTUAM-16-12, IPPP-16-27, DCTP-16-54, KCL-PH-TH-2016-19",
    doi = "10.1007/JHEP09(2016)048",
    journal = "JHEP",
    volume = "05",
    pages = "118",
    year = "2016",
    note = "[Erratum: JHEP 09, 048 (2016)]"
}

@article{Boehm:2018sux,
    author = "B\oe{}hm, C. and Cerde\~no, D. G. and Machado, P. A. N. and Olivares-Del Campo, A. and Perdomo, E. and Reid, E.",
    title = "{How high is the neutrino floor?}",
    eprint = "1809.06385",
    archivePrefix = "arXiv",
    primaryClass = "hep-ph",
    reportNumber = "IPPP/18/72, FERMILAB-PUB-18-486-T, IPPP/18/72; FERMILAB-PUB-18-486-T",
    doi = "10.1088/1475-7516/2019/01/043",
    journal = "JCAP",
    volume = "01",
    pages = "043",
    year = "2019"
}

@article{Vergados:2008jp,
    author = "Vergados, J. D. and Ejiri, H.",
    title = "{Can Solar Neutrinos be a Serious Background in Direct Dark Matter Searches?}",
    eprint = "0805.2583",
    archivePrefix = "arXiv",
    primaryClass = "hep-ph",
    doi = "10.1016/j.nuclphysb.2008.06.004",
    journal = "Nucl. Phys. B",
    volume = "804",
    pages = "144--159",
    year = "2008"
}

@article{Monroe:2007xp,
    author = "Monroe, Jocelyn and Fisher, Peter",
    title = "{Neutrino Backgrounds to Dark Matter Searches}",
    eprint = "0706.3019",
    archivePrefix = "arXiv",
    primaryClass = "astro-ph",
    doi = "10.1103/PhysRevD.76.033007",
    journal = "Phys. Rev. D",
    volume = "76",
    pages = "033007",
    year = "2007"
}

@article{Pospelov:2011ha,
    author = "Pospelov, Maxim",
    title = "{Neutrino Physics with Dark Matter Experiments and the Signature of New Baryonic Neutral Currents}",
    eprint = "1103.3261",
    archivePrefix = "arXiv",
    primaryClass = "hep-ph",
    doi = "10.1103/PhysRevD.84.085008",
    journal = "Phys. Rev. D",
    volume = "84",
    pages = "085008",
    year = "2011"
}

@article{Gutlein:2010tq,
    author = "Gutlein, A. and others",
    title = "{Solar and atmospheric neutrinos: Background sources for the direct dark matter search}",
    eprint = "1003.5530",
    archivePrefix = "arXiv",
    primaryClass = "hep-ph",
    doi = "10.1016/j.astropartphys.2010.06.002",
    journal = "Astropart. Phys.",
    volume = "34",
    pages = "90--96",
    year = "2010"
}

@article{PandaX-II:2021nsg,
    author = "Cheng, Chen and others",
    collaboration = "PandaX-II",
    title = "{Search for Light Dark Matter-Electron Scatterings in the PandaX-II Experiment}",
    eprint = "2101.07479",
    archivePrefix = "arXiv",
    primaryClass = "hep-ex",
    doi = "10.1103/PhysRevLett.126.211803",
    journal = "Phys. Rev. Lett.",
    volume = "126",
    number = "21",
    pages = "211803",
    year = "2021"
}

@inproceedings{Essig:2022dfa,
    author = "Essig, Rouven and others",
    title = "{Snowmass2021 Cosmic Frontier: The landscape of low-threshold dark matter direct detection in the next decade}",
    booktitle = "{Snowmass 2021}",
    eprint = "2203.08297",
    archivePrefix = "arXiv",
    primaryClass = "hep-ph",
    reportNumber = "FERMILAB-CONF-22-181-PPD",
    month = "3",
    year = "2022"
}

@article{Khan:2019jvr,
    author = "Khan, Amir N. and Rodejohann, Werner and Xu, Xun-Jie",
    title = "{Borexino and general neutrino interactions}",
    eprint = "1906.12102",
    archivePrefix = "arXiv",
    primaryClass = "hep-ph",
    reportNumber = "FERMILAB-PUB-19-348-T",
    doi = "10.1103/PhysRevD.101.055047",
    journal = "Phys. Rev. D",
    volume = "101",
    number = "5",
    pages = "055047",
    year = "2020"
}

@article{Bilmis:2015lja,
    author = "Bilmis, S. and Turan, I. and Aliev, T. M. and Deniz, M. and Singh, L. and Wong, H. T.",
    title = "{Constraints on Dark Photon from Neutrino-Electron Scattering Experiments}",
    eprint = "1502.07763",
    archivePrefix = "arXiv",
    primaryClass = "hep-ph",
    doi = "10.1103/PhysRevD.92.033009",
    journal = "Phys. Rev. D",
    volume = "92",
    number = "3",
    pages = "033009",
    year = "2015"
}

@article{Agarwalla:2012wf,
    author = "Agarwalla, Sanjib Kumar and Lombardi, Francesco and Takeuchi, Tatsu",
    title = "{Constraining Non-Standard Interactions of the Neutrino with Borexino}",
    eprint = "1207.3492",
    archivePrefix = "arXiv",
    primaryClass = "hep-ph",
    reportNumber = "EURONU-WP6-12-52, IFIC-12-49",
    doi = "10.1007/JHEP12(2012)079",
    journal = "JHEP",
    volume = "12",
    pages = "079",
    year = "2012"
}

@article{Nieves,
  title = {Electromagnetic properties of Majorana neutrinos},
  author = {Nieves, Jos\'e F.},
  journal = {Phys. Rev. D},
  volume = {26},
  issue = {11},
  pages = {3152--3158},
  numpages = {0},
  year = {1982},
  month = {Dec},
  publisher = {American Physical Society},
  doi = {10.1103/PhysRevD.26.3152},
  url = {https://link.aps.org/doi/10.1103/PhysRevD.26.3152}
}

@article{Kayser,
  title = {Majorana neutrinos and their electromagnetic properties},
  author = {Kayser, Boris},
  journal = {Phys. Rev. D},
  volume = {26},
  issue = {7},
  pages = {1662--1670},
  numpages = {0},
  year = {1982},
  month = {Oct},
  publisher = {American Physical Society},
  doi = {10.1103/PhysRevD.26.1662},
  url = {https://link.aps.org/doi/10.1103/PhysRevD.26.1662}
}

@article{Cabral-Rosetti:2002zyl,
    author = "Cabral-Rosetti, L. G. and Moreno, M. and Rosado, A.",
    editor = {Diaz-Cruz, J. Lorenzo and Engelfried, J\"urgen and Kirchbach, Mariana and Mondragon, Myriam},
    title = "{Dirac neutrino anapole moment}",
    eprint = "hep-ph/0206083",
    archivePrefix = "arXiv",
    doi = "10.1063/1.1489777",
    journal = "AIP Conf. Proc.",
    volume = "623",
    number = "1",
    pages = "347--350",
    year = "2002"
}

@article{migdal:1939svj,
    author = "A. Migdal",
    title = "{Ionizatsiya atomov pri yadernykh reaktsiyakh}",
    journal = "Sov. Phys. JETP ",
    volume = "9",
    pages = "1163",
    year = "1939"
}

@article{Evans:2018bqy,
    author = "Evans, N. Wyn and O'Hare, Ciaran A. J. and McCabe, Christopher",
    title = "{Refinement of the standard halo model for dark matter searches in light of the Gaia Sausage}",
    eprint = "1810.11468",
    archivePrefix = "arXiv",
    primaryClass = "astro-ph.GA",
    doi = "10.1103/PhysRevD.99.023012",
    journal = "Phys. Rev. D",
    volume = "99",
    number = "2",
    pages = "023012",
    year = "2019"
}

@article{Bozorgnia:2016ogo,
    author = "Bozorgnia, Nassim and Calore, Francesca and Schaller, Matthieu and Lovell, Mark and Bertone, Gianfranco and Frenk, Carlos S. and Crain, Robert A. and Navarro, Julio F. and Schaye, Joop and Theuns, Tom",
    title = "{Simulated Milky Way analogues: implications for dark matter direct searches}",
    eprint = "1601.04707",
    archivePrefix = "arXiv",
    primaryClass = "astro-ph.CO",
    doi = "10.1088/1475-7516/2016/05/024",
    journal = "JCAP",
    volume = "05",
    pages = "024",
    year = "2016"
}

@article{Necib:2018igl,
    author = "Necib, Lina and Lisanti, Mariangela and Garrison-Kimmel, Shea and Wetzel, Andrew and Sanderson, Robyn and Hopkins, Philip F. and Faucher-Gigu\`ere, Claude-Andr\'e and Kere\v{s}, Du\v{s}an",
    title = "{Under the Firelight: Stellar Tracers of the Local Dark Matter Velocity Distribution in the Milky Way}",
    eprint = "1810.12301",
    archivePrefix = "arXiv",
    primaryClass = "astro-ph.GA",
    doi = "10.3847/1538-4357/ab3afc",
    month = "10",
    year = "2018"
}

@article{Vitagliano:2019yzm,
    author = "Vitagliano, Edoardo and Tamborra, Irene and Raffelt, Georg",
    title = "{Grand Unified Neutrino Spectrum at Earth: Sources and Spectral Components}",
    eprint = "1910.11878",
    archivePrefix = "arXiv",
    primaryClass = "astro-ph.HE",
    reportNumber = "MPP-2019-205",
    doi = "10.1103/RevModPhys.92.045006",
    journal = "Rev. Mod. Phys.",
    volume = "92",
    pages = "45006",
    year = "2020"
}

@article{Mohapatra,
  title = {Local $B\ensuremath{-}L$ Symmetry of Electroweak Interactions, Majorana Neutrinos, and Neutron Oscillations},
  author = {Mohapatra, R. N. and Marshak, R. E.},
  journal = {Phys. Rev. Lett.},
  volume = {44},
  issue = {20},
  pages = {1316--1319},
  numpages = {0},
  year = {1980},
  month = {May},
  publisher = {American Physical Society},
  doi = {10.1103/PhysRevLett.44.1316},
  url = {https://link.aps.org/doi/10.1103/PhysRevLett.44.1316}
}

@article{Vogel:1989iv,
    author = "Vogel, P. and Engel, J.",
    title = "{Neutrino Electromagnetic Form-Factors}",
    reportNumber = "CALT-63-539",
    doi = "10.1103/PhysRevD.39.3378",
    journal = "Phys. Rev. D",
    volume = "39",
    pages = "3378",
    year = "1989"
}

@article{Besla:2019xbx,
    author = "Besla, Gurtina and Peter, Annika and Garavito-Camargo, Nicolas",
    title = "{The highest-speed local dark matter particles come from the Large Magellanic Cloud}",
    eprint = "1909.04140",
    archivePrefix = "arXiv",
    primaryClass = "astro-ph.GA",
    doi = "10.1088/1475-7516/2019/11/013",
    journal = "JCAP",
    volume = "11",
    pages = "013",
    year = "2019"
}

@article{davidson,
  title = {$B\ensuremath{-}L$ as the fourth color within an $\mathrm{SU}{(2)}_{L}\ifmmode\times\else\texttimes\fi{}\mathrm{U}{(1)}_{R}\ifmmode\times\else\texttimes\fi{}\mathrm{U}(1)$ model},
  author = {Davidson, Aharon},
  journal = {Phys. Rev. D},
  volume = {20},
  issue = {3},
  pages = {776--783},
  numpages = {0},
  year = {1979},
  month = {Aug},
  publisher = {American Physical Society},
  doi = {10.1103/PhysRevD.20.776},
  url = {https://link.aps.org/doi/10.1103/PhysRevD.20.776}
}

@article{AtzoriCorona:2023ais,
    author = "Atzori Corona, M. and Cadeddu, M. and Cargioli, N. and Dordei, F. and Giunti, C.",
    title = "{On the impact of the Migdal effect in reactor CE$\nu$NS experiments}",
    eprint = "2307.12911",
    archivePrefix = "arXiv",
    primaryClass = "hep-ph",
    month = "7",
    year = "2023"
}

@article{AtzoriCorona:2022jeb,
    author = "Atzori Corona, M. and Bonivento, W. M. and Cadeddu, M. and Cargioli, N. and Dordei, F.",
    title = "{New constraint on neutrino magnetic moment and neutrino millicharge from LUX-ZEPLIN dark matter search results}",
    eprint = "2207.05036",
    archivePrefix = "arXiv",
    primaryClass = "hep-ph",
    doi = "10.1103/PhysRevD.107.053001",
    journal = "Phys. Rev. D",
    volume = "107",
    number = "5",
    pages = "053001",
    year = "2023"
}

@article{XENON:2022ltv,
    author = "Aprile, E. and others",
    collaboration = "XENON",
    title = "{Search for New Physics in Electronic Recoil Data from XENONnT}",
    eprint = "2207.11330",
    archivePrefix = "arXiv",
    primaryClass = "hep-ex",
    doi = "10.1103/PhysRevLett.129.161805",
    journal = "Phys. Rev. Lett.",
    volume = "129",
    number = "16",
    pages = "161805",
    year = "2022"
}

@article{GrillidiCortona:2020owp,
    author = "Grilli di Cortona, Giovanni and Messina, Andrea and Piacentini, Stefano",
    title = "{Migdal effect and photon Bremsstrahlung: improving the sensitivity to light dark matter of liquid argon experiments}",
    eprint = "2006.02453",
    archivePrefix = "arXiv",
    primaryClass = "hep-ph",
    doi = "10.1007/JHEP11(2020)034",
    journal = "JHEP",
    volume = "11",
    pages = "034",
    year = "2020"
}

@article{Demirci:2023tui,
    author = "Demirci, Mehmet and Mustamin, M. Fauzi",
    title = "{Solar neutrino constraints on light mediators through coherent elastic neutrino-nucleus scattering}",
    eprint = "2312.17502",
    archivePrefix = "arXiv",
    primaryClass = "hep-ph",
    doi = "10.1103/PhysRevD.109.015021",
    journal = "Phys. Rev. D",
    volume = "109",
    number = "1",
    pages = "015021",
    year = "2024"
}

@article{DeRomeri:2024dbv,
    author = "De Romeri, Valentina and Papoulias, Dimitrios K. and Ternes, Christoph A.",
    title = "{Light vector mediators at direct detection experiments}",
    eprint = "2402.05506",
    archivePrefix = "arXiv",
    primaryClass = "hep-ph",
    month = "2",
    year = "2024"
}

@article{Lindner:2024eng,
    author = "Lindner, Manfred and Rink, Thomas and Sen, Manibrata",
    title = "{Light vector bosons and the weak mixing angle in the light of new reactor-based CE$\nu$NS experiments}",
    eprint = "2401.13025",
    archivePrefix = "arXiv",
    primaryClass = "hep-ph",
    month = "1",
    year = "2024"
}

@article{Schwemberger:2023hee,
    author = "Schwemberger, Thomas and Takhistov, Volodymyr and Yu, Tien-Tien",
    title = "{Hunting Nonstandard Neutrino Interactions and Leptoquarks in Dark Matter Experiments}",
    eprint = "2307.15736",
    archivePrefix = "arXiv",
    primaryClass = "hep-ph",
    reportNumber = "KEK-QUP-2023-0007, KEK-TH-2516, KEK-Cosmo-0309, IPMU23-0009",
    month = "7",
    year = "2023"
}
\end{document}